# The Strategic Reference Gene: an organismal theory of inclusive fitness

**Authors:** Lutz Fromhage[1]*, Michael D Jennions[2]

**Affiliations:**

[1] *Department of Biological and Environmental Science, University of Jyvaskyla, P.O. Box 35, 40014 Jyvaskyla, Finland*

[2] *Ecology & Evolution, Research School of Biology, The Australian National University, Canberra ACT 2601, Australia*

*Correspondence to: lutz.fromhage@jyu.fi

How to define and use the concept of inclusive fitness is a contentious topic in evolutionary theory. Inclusive fitness can be used to calculate selection on a focal *gene*, but it is also applied to whole *organisms*. Individuals are then predicted to appear designed as if to maximise their inclusive fitness, provided that certain conditions are met (formally when interactions between individuals are 'additive'). Here we argue that applying the concept of inclusive fitness to organisms is justified under far broader conditions than previously shown, but only if it is appropriately defined. Specifically, we propose that organisms should maximise the sum of their offspring (*including* any accrued due to the behaviour/phenotype of relatives), plus any effects on their relatives' offspring production, weighted by relatedness. In contrast, most theoreticians have argued that a focal individual's inclusive fitness should exclude any offspring accrued due to the behaviour of relatives. Our approach is based on the notion that long-term evolution follows the genome's 'majority interest' of building coherent bodies that are efficient 'vehicles' for gene propagation. A gene favoured by selection that reduces the propagation of unlinked genes at other loci (e.g. meiotic segregation distorters that lower sperm production) is eventually neutralised by counter-selection throughout the rest of the genome. Most phenotypes will therefore appear as if designed to maximise the propagation of any given gene in a focal individual and its relatives.

*Perhaps we should not feel entirely confident about generalizing our principle until a more comprehensive mathematical argument, with inclusive fitness more widely defined, has been worked out.* – Hamilton [1]

## 1. Introduction

What, if anything, are organisms shaped by evolution adapted to achieve [2–4]? To answer this question, consider the fact that natural selection is roughly analogous to trial-and-error learning: mutations create gene variants which affect the phenotypes of organisms expressing them; variants then spread if their causal effects on the world, mediated by how they affect the phenotype, aid their propagation [5]. Accordingly, it is a truism that any naturally selected trait can be said to have evolved because genes contributing to the trait in past generations were more successful than their alternatives at leaving copies in the present. But what kinds of phenotypes will successful genes contribute to build? Hamilton made a major breakthrough in answering this question [6,7]. He distinguished two causal pathways by which a gene, expressed in a given organism, can aid its propagation. It can enhance the organism's own reproduction (direct fitness), and it can cause the organism to enhance the reproduction of others that carry the gene's identical copies (indirect fitness). To capture this insight, he defined *inclusive fitness* ($IF_{\mathrm{Hamilton}}$) as a combined measure of direct and indirect fitness components [6]:
"Inclusive fitness may be imagined as the personal fitness which an individual actually expresses in its production of adult offspring as it becomes after it has been first stripped and then augmented in a certain way. It is stripped of all components which can be considered as due to the individual's social environment, leaving the fitness which he would express if not exposed to any of the harms or benefits of that environment. This quantity is then augmented by certain fractions of the quantities of harm and benefit which the individual himself causes to the fitnesses of his neighbours. The fractions in question are simply the coefficients of relationship appropriate to the neighbours whom he affects: unity for clonal individuals, one-half for sibs, one-quarter for half-sibs, one-eighth for cousins, ... and finally zero for all neighbours whose relationship can be considered negligibly small".

Hamilton showed that $IF_{\mathrm{Hamilton}}$ works as a *genetic accounting tool* to predict when a focal gene is positively selected, which occurs when an individual expressing it enjoys increased inclusive fitness. He inferred from this that the long-term outcome of successive genes being selected in this way is that organisms shaped by natural selection should be adapted to maximise $IF_{\mathrm{Hamilton}}$. This would make $IF_{\mathrm{Hamilton}}$ a *phenotypic maximand* [3]. The concept of a phenotypic maximand is useful for studying adaptation because we can then envisage individual organisms as maximising agents with a defined biological purpose [8]. It allows us to predict that an organism's (naturally selected) traits tend to be shaped to cause a higher expected value of the maximand than feasible alternative traits. Organism-centred usage of inclusive fitness requires that $IF_{\mathrm{Hamilton}}$ is a measurable property of an individual organism. To meet this requirement, $IF_{\mathrm{Hamilton}}$ must use a concept of relatedness that is applicable to entire organisms (i.e., that approximately measures genetic similarity across the genome), rather than being only applicable to one gene at a time.

By contrast, inclusive fitness models often focus on a single gene, predicting that it will spread if it satisfies Hamilton's rule $rb - c > 0$ (where $r$ is relatedness, $-c$ and $b$ are changes caused to the reproduction of 'self' and 'other', and the left-hand-side is defined as the gene's *inclusive fitness effect* [6,9]). This approach calls for a gene-specific ('genic') definition of relatedness [10] which – unlike 'pedigree relatedness' between organisms – accounts for genetic similarity between individuals for a focal gene that can arise by processes that do not apply equally to all genes (e.g. non-random assortment of organisms with the focal gene). This difference in relatedness concepts indicates that the connection between gene-level selection and organism-level adaptation is not straightforward. Indeed, some theorists have even concluded that inclusive fitness is not a meaningful property of an organism [11–13]. If true, this precludes it being a phenotypic maximand (but see [3,9]). But do we really want to abandon the use of inclusive fitness when we study adaptations, which are usually complex traits determined by the effects of many genes?

Here we argue that invoking $IF_{\mathrm{Hamilton}}$ as a general phenotypic maximand is problematic, but that these problems are surmounted if we redefine inclusive fitness. We start from the observation that genes with opposing phenotypic effects can simultaneously be selected for, due to gene-specific patterns of inheritance and expression (e.g. meiotic driver genes versus those for balanced meiosis). We then invoke a broad interpretation of the principle of the 'parliament of genes' [14] to predict how such opposing forces are likely to be resolved over evolutionary time. To operationally characterise the genome's 'majority interest', we invoke an idealised 'reference gene' whose interest in which phenotype is expressed always aligns with that of most other genes in the same organism. We then propose a modified definition of inclusive fitness based on a quantity whose maximisation best serves the genome's 'majority interest'. Our goal is not to paint a precise picture of population genetic processes, but rather to argue for a higher-level principle that tends to guide cumulative phenotypic evolution in a coherent direction: namely, towards optimised design of individual organisms.

We consider a wide range of potential objections to our approach, which is likely to be controversial. However, to avoid too many asides, we relegate many of these objections to a 'Questions and Answers' list (Supplementary Material). We also include a video that gives a non-technical overview of our ideas.

## 2. Reference genes and the parliament of genes

Any quantity that qualifies as a phenotypic maximand should tend to be increased through phenotypic changes induced by gene frequency changes due to natural selection. But, of course, organisms are integrated units shaped by selection on thousands of loci over long timespans, so not every positively selected gene needs to be a step towards increasing the maximand. Once a focal gene has spread and propelled a population along an evolutionary trajectory in phenotypic space, genetic variation at other loci determines how the trajectory continues. The focal gene's contribution could either be retained or eliminated. When studying long-term evolution, the common guiding question: "what kind of gene will be positively selected?" should therefore be complemented by adding "…, such that its phenotypic effect is not eliminated in the long run". A similar point was made by Leigh [14] to account for fair meiosis being overwhelmingly common, despite the huge selective advantage that segregation distorter genes can enjoy. Leigh wrote: "It is as if we had to do with a parliament of genes: each acts in its own self-interest, but if its acts hurt others, they will combine together to suppress it." He explained the rarity of segregation distorters by invoking the principle that genes that oppose the genome's 'majority interest' are eliminated by counter-selection at other loci. Here we combine this idea with Dawkins' [4] vision of individual organisms as *vehicles* for gene propagation. Specifically, we postulate that the genome's 'majority interest' is to build an organism with high *vehicle quality*, which we define as an organism's general capacity to propagate its genes and their identical copies. To quantify vehicle quality we envisage a hypothetical *reference gene* (more precisely, an allele) which is: (i) present in the focal organism, (ii) rare in the population (iii) subject to Mendelian inheritance, and (iv) rarely or never expressed (i.e. low penetrance; assuming that other alleles at the same locus are never expressed). These properties are chosen in part (i, ii) to facilitate measuring gene propagation (essentially, by counting copies), and in part (iii, iv) so that the reference gene's evolutionary interest as to what phenotype should be expressed (i.e. the ranking of possible phenotypes with respect to how well they propagate the reference gene) aligns with the common interest of the organism's other genes.

We measure vehicle quality as the number of reference gene copies that can be causally attributed to the focal organism. These are the net number of additional copies that arise because the focal organism exists. Every sexually produced offspring of the focal organism contributes $s$ copies, and every offspring produced by a relative of degree $r$ accounts for $sr$ copies. Here, $s$ is the probability of transmitting a reference gene copy to a given offspring, which is given by the focal organism's consanguinity [15] with itself; in diploid, outbreeding populations, $s = 0.5$. The pedigree relatedness $r$ is the coefficient of relatedness [15] as applied to weakly selected genes due to coancestry. The number of propagated reference gene copies then sums to $s \cdot \sum(r \cdot \Delta n_r)$, where $\Delta n_r$ is the net number of offspring[1] produced (or not produced) by relatives of degree $r$ because the focal organism exists. It includes *all* of the focal organism's own offspring, for which $r = 1$. An individual's vehicle quality is maximised by the phenotype that causes the greatest representation of the reference gene in future generations. This occurs when the individual maximises the expected value of $\sum(r \cdot \Delta n_r)$, which is the sum of its own offspring number, plus its effects on its relatives' number of offspring, weighted by relatedness. We call this the *folk definition of inclusive fitness*, $IF_{\text{folk}}$, which has been described as "a common misdefinition of inclusive fitness" [16]. Unlike $IF_{\text{Hamilton}}$ there is no 'stripping' of the social environment for $IF_{\text{folk}}$. That is, all of the focal individual's own offspring count as being caused by it, in the sense that they would not have been produced if the focal organism hadn't existed and exhibited a phenotype with the requisite fertility.

[1] To be exact, to quantify each reference gene copy's projected contribution to the future gene pool, each offspring should be weighted by $V/l$, where reproductive value $V$ is the offspring's projected contribution to the future gene pool, and ploidy level $l$ accounts for the fact that a diploid offspring's contribution is shared between two haploid genomes.

To summarise, we postulate that the genome's 'majority interest' is to build an organism with high vehicle quality. Here, vehicle quality is the general capacity for gene propagation, which we propose to quantify as the number of reference gene copies that can be causally attributed to the organism. Since that number is proportional to $IF_{\text{folk}}$, the number of reference gene copies is maximised when $IF_{\text{folk}}$ is maximised. So, if evolution mainly follows the genome's majority interest, organisms should express traits that maximise their $IF_{\text{folk}}$. The reference gene's property of being rarely expressed (hence weakly selected) justifies using a pedigree-based concept of relatedness for $IF_{\text{folk}}$, which is also relevant for multi-locus evolution because coancestry is the only source of genetic similarity that promotes wide agreement across the genome as to what traits best serve each constituent gene's propagation [3,10].

What do we mean when we claim that organisms should behave so as to maximise their $IF_{\text{folk}}$? In general, maximization occurs when a mathematical or physical function reaches its highest achievable output value through changes, within a specified range, in the values of its input arguments. In the present case, the function of interest is $IF_{\text{folk}}$, and its argument is the individual organism's phenotypic strategy (including its propensity to help or harm, but also non-social traits). Formally we can write this as $IF_{\text{folk}}[\pi] = \sum(r \cdot \Delta n_r)\,|do(phenotype = \pi)$, where the 'do' operator (adopted from Pearl's causal modelling framework [17,18]) stands for 'set phenotype to $\pi$'. This formulation conveys the idea that any given phenotype belongs to a set of feasible options that could be generated by appropriate genotypes, or by experimental intervention. We then predict that organisms tend to exhibit phenotypes that yield higher $IF_{\text{folk}}$ than feasible alternatives. Crucially, while $IF_{\text{folk}}$ is useful for comparing phenotypes at a given time, in a given social environment, it does not measure changes in absolute fit between organisms and their environment over evolutionary time. Hence our prediction that phenotypes yielding higher $IF_{\text{folk}}$ tend to evolve should not be misinterpreted as a claim that $IF_{\text{folk}}$ increases over evolutionary time, towards a maximum at equilibrium. The environment that sets the background for evaluating $IF_{\text{folk}}$ changes over time due to both abiotic and biotic factors, including frequency-dependent traits.

## 3. Rogue genes

Despite the parliament of genes, selection need not always increase vehicle quality. At least in the short term, the opposite can occur. Here we use the term *rogue genes* for genes that can generate selection for traits that reduce vehicle quality. Rogue genes include *Mendelian outlaw genes*, *greenbeard genes*, and a previously undescribed type that we call *mirror effect rogue genes.* The existence of these kinds of genes is partly why some theoreticians are dubious about the usefulness of applying inclusive fitness to individual organisms. *Mendelian outlaw genes* spread at the expense of unlinked genes in the same organism by violating the laws of Mendelian inheritance. A meiotic drive gene that ends up in more than half of an organism's zygotes may spread, despite reducing the organism's reproductive output. However, a driver gene also selects for unlinked modifier genes that neutralise its phenotypic effect [19]. *Greenbeard genes* can spread by causing their bearer to (i) exhibit an cue (e.g. a green beard), and (ii) behave altruistically towards others bearing the cue [4,20]. Once a greenbeard gene has spread, the maintenance of its phenotypic effects relies on the genetic constraint that the cue (which enhances vehicle quality) cannot be expressed without the altruistic behaviour (which reduces vehicle quality). Eventually this constraint should be undermined through selection for modifier genes that suppress the altruistic behaviour, but not the cue [21,22]. *Mirror effect rogue genes* are particularly pertinent to deciding whether $IF_{\text{folk}}$ qualifies as a phenotypic maximand, but we defer their definition until Section 5 as we must first introduce some additional concepts.

## 4. The mirror effect

There is a conceptual distinction between genes with and without a 'mirror effect'. The 'mirror effect' is a gene's tendency to be simultaneously expressed in interacting individuals that carry the gene. The term alludes to the idea that an individual expressing a gene with a mirror effect will tend to find its own phenotype 'mirrored' by relatives who share the gene. In an interaction between individuals who share a gene, the mirror effect's strength is quantified as the conditional probability that the gene is expressed in the non-focal individual, given that it is expressed in the focal individual. When this probability is zero or negligibly small, we speak of a 'gene without mirror effect'. Population genetic models (including Hamilton's [6]) often assume that a gene is always expressed, thereby implicitly assuming the mirror effect is maximally strong. There are, however, two mechanisms by which a gene can be exempt from the mirror effect. First, if the expression of a behaviour is conditional on an asymmetry between social partners (e.g. in size, residency, caste, social dominance, or any arbitrary variable), the underlying gene is exempt from the mirror effect [23]. Second, if a gene has low penetrance (i.e., probability of being expressed) it will rarely be simultaneously expressed in both the actor and the recipient during a social interaction – even if both parties carry the gene. This makes the mirror effect negligibly weak. The mirror effect presents a difficulty for quantifying the causal effects of a gene because it is expressed both in a focal organism and in other organisms that make up its social environment (Fig. 1). For example, if we compare organisms that either do or do not have a helping gene (with mirror effect), $IF_{\text{folk}}$ overestimates the gene's causal effect because it counts the benefit of helping twice – both when giving and receiving help [16]. The conventional remedy for this 'double accounting' is to use $IF_{\text{Hamilton}}$, which, by 'stripping' the effect of the social environment, isolates the causal effect of a gene when it is expressed in the focal organism. However, inspired by Pearl's causal modelling framework [17,18], we suggest an alternative remedy that is analogous to measuring causality in a controlled experiment. We can measure a gene's causal effect on a focal organism's $IF_{\text{folk}}$ by comparing the observed value of $IF_{\text{folk}}$ with the counterfactual value $\widehat{IF}_{\text{folk}}$ that would arise if the individual were experimentally prevented from expressing the gene (see legend to Fig. 1). This heuristic recovers the correct inclusive fitness effect when interactions are additive (i.e., when the effects of an individual's actions are independent of the phenotype of others; Fig.1). As importantly, it also predicts the direction of multi-locus evolution for the kinds of non-additive interactions that have stymied attempts to 'strip' the effects of the social environment on the focal individual's inclusive fitness (Supplementary Material 5, Q15). Instead of being a mere technicality that needs accounting for, the mirror effect can sometimes affect the direction of selection by biasing the flow of social benefits towards particular genotypes in non-additive interactions (i.e., when the benefits provided to a recipient partly depend on the recipient's phenotype; Fig. 2).

## 5. Mirror effect rogue genes

Intriguingly, opposite phenotypes (e.g. help versus do not help) can be selected for depending on whether or not a gene has a mirror effect (Fig. 2). In this context we define a *mirror effect rogue (MER) gene* as an allele that reduces the *vehicle quality* of the organisms expressing it, but is still selected for due to the mirror effect (i.e. because the mirror effect biases the flow of social benefits towards particular genotypes at that locus). Here, an organism's reduction in vehicle quality is measured relative to the counterfactual situation where only the focal organism, in its given social environment, expresses an alternative phenotype to that induced by the MER. This definition implies that any unlinked modifier gene will be selected for if it slightly reduces a MER gene's probability of being expressed. This follows because the modifier gene meets our definition of a reference gene in being rarely expressed (only in rare instances where its effect on the MER gene is realised), implying that more copies of it are propagated when the focal organism's vehicle quality is increased (due to the MER gene's negative effect being negated by the modifier). MER genes can occur when there are non-additive social interactions in which matching phenotypes interfere with each other (e.g. mutual help is less efficient than unilateral help; Supplementary material 1). Loosely speaking, these are conditions where a rational actor would prefer to help, unless she anticipates that her actions will be 'mirrored' by relatives. In the example given in Fig 2, helping increases vehicle quality when it is rare; however, a MER allele for 'not helping' can spread to fixation when the helping allele is always expressed (i.e. is subject to a mirror effect), thereby failing to generate indirect fitness benefits for its carriers due to interference. We should emphasize that MER genes do not merit discussion because there is empirical evidence for them, but rather because many theoretical models [24–27] have made assumptions under which MER genes occur. This has prompted conclusions which appear to contradict our prediction that evolution tends toward the maximisation of $IF_{\mathrm{folk}}$.

**6. The folk definition of inclusive fitness**

Based on our definition of vehicle quality and $IF_{\mathrm{folk}}$ we advance a heuristic argument about cumulative change, and a deductive argument about evolutionary stability, to infer the most likely outcome of long-term natural selection. Consider a positively selected focal gene (of any effect size, hence subject to any strength of selection) for a trait that increases vehicle quality through an initially inefficient mechanism, as is likely for novel traits. Other genes elsewhere in the genome that enhance the trait's efficiency will then increase vehicle quality further and be selected for. In this way, traits that increase vehicle quality have the potential to evolve through complementary, cumulative contributions from unlinked genes. This potential is crucial because many genes (with various effect sizes) are usually involved in producing finely adapted and/or complex traits. It is exceedingly rare for such traits to arise in a single mutational step. Conversely, if a focal gene promotes development of a trait that reduces vehicle quality while facilitating its own propagation (i.e., a rogue gene), the trait faces counter-selection from elsewhere in the genome. The likely success of the 'parliament of genes' in countering a rogue gene is aided by the architectural principle that complex structures are more easily destroyed than built. For example, if trait development depends on a suite of genes that interact in a coherent fashion, then mutations disrupting any of these myriad interactions will tend to derail its development. These twin considerations suggest that traits that increase vehicle quality will prevail in the long run, even if selection for rogue genes temporarily reverses the trend.

We next make an argument about evolutionary stability. Consider a mutant gene whose expression in a focal individual induces a phenotypic change that increases the individual's $IF_{\mathrm{folk}}$. If this gene meets our definition of a reference gene, it is guaranteed to be positively selected because $IF_{\mathrm{folk}}$ is defined by a reference gene's propagation success. Hence, no phenotypic strategy is evolutionarily stable unless the organisms adopting it already maximize their $IF_{\mathrm{folk}}$. To reach this conclusion, all we need to assume is that mutations can arise with any degree of penetrance. Even if evolutionary dynamics are largely driven by high-penetrance genes under strong selection, evolutionary stability has to be evaluated allowing for mutant genes with any degree of penetrance. To the extent that the availability of suitable alleles poses a genetic constraint, even a low frequency of mutations should eventually overcome this constraint. Hammerstein [28] made a similar point about non-social evolution: "If genetic constraints keep a population away from a phenotypically adaptive state, there is a possibility for a new mutant allele to code for phenotypes that perform better than the population mean." It follows that the maximisation of $IF_{\mathrm{folk}}$ is necessary for evolutionary stability under far broader conditions than have been previously reported [27], including non-additive interactions and mutations of various step sizes, both large and small.

We emphasise that the above argument neither assumes nor implies that low-penetrance genes are more important for evolutionary stability than high-penetrance genes. However, it is stability against low-penetrance mutations that implies organismal maximising behaviour. This is because a low-penetrance gene, when expressed, induces exactly the kind of change we envision in our definition of $IF_{\text{folk}}$ being a function of phenotype: namely, there is a change in the focal organism but no correlated (mirrored) change in its social environment. The gene's causal effect on its own propagation thus corresponds exactly to its causal effect on the focal organism's $IF_{\text{folk}}$. And this correspondence ensures that only organisms that already maximise their $IF_{\text{folk}}$ cannot be modified by a low-penetrance gene to gain a propagation advantage.

Although necessary, maximisation of $IF_{\text{folk}}$ is not sufficient for evolutionary stability. Even when $IF_{\text{folk}}$ is maximised and it cannot be increased by changing a focal organism's phenotype in its current environment, a large-effect mutation with mirror effect might perturb the social environment so as to render a new phenotype optimal. For example, if there are synergistic benefits of mutual cooperation, cooperator genes with mirror effect can invade (and then increase $IF_{\text{folk}}$ in the new local environment they create) even when unilateral switching to cooperation would decrease $IF_{\text{folk}}$ (Supplementary Material 1).

Earlier work that rejected the principle of $IF_{\text{folk}}$ maximisation made the restrictive assumption that genes with incomplete penetrance and/or conditional expression do not exist [24,25]. Consequently, mutant genes could not change the phenotype of the individual they were expressed in without immediately facing a correlated change in relatives carrying the same gene. Given interference between matching phenotypes, which is when MER genes can arise, this prevented organisms from evolving the optimal phenotype for their social environment (see Figure 2). Here we show that equilibria established by MER genes (at which $IF_{\text{folk}}$ is not maximised) are unstable against invasion by mutant genes without mirror effect, whereas the corresponding equilibria at which $IF_{\text{folk}}$ is maximised are stable against mutant genes both with and without mirror effect (Supplementary Material 1). We then use simulations to show that the principle of $IF_{\text{folk}}$ maximisation is realised ever more closely when the genetic system is more flexible (Supplementary Material 2). This flexibility can arise due to either a one-locus multi-allele system (Figure S1) or a multi-locus system (Figures S2-S4). Our results suggest that, barring permanent genetic constraints that seem biologically implausible, interference between matching phenotypes (that allows for MER genes) poses no unsurmountable impediment to organisms evolving the optimal phenotype for their environment in the long-term.

## 7. Hamilton's inclusive fitness

Does maximising $IF_{\text{folk}}$ instead of $IF_{\text{Hamilton}}$ actually makes a difference? Do we really need to abandon $IF_{\text{Hamilton}}$? To be a quantity which an individual could meaningfully be said to be maximising, $IF_{\text{Hamilton}}$, like $IF_{\text{folk}}$, must be a function of an individual organism's phenotype. This raises the question of how to interpret the 'stripping procedure' in Hamilton's definition. Hamilton stated that $IF_{\text{Hamilton}}$ is "*stripped of all components which can be considered as due to* [i.e., that are causal effects of] *the individual's social environment, leaving the fitness which he would express if not exposed to any of the harms or benefits of that environment*." We take this to mean that, if a non-focal individual performs a social act that causes the focal organism's reproduction to change (compared to the counterfactual situation where it is not performed), then the magnitude of that change must be stripped from the focal individual's *IF*. This worked in Hamilton's original setup because the assumption of additive interactions ensures that every consequence is attributable to a single act and actor. Additivity ensures that the components to be stripped are unaffected by the focal organism. By contrast, non-additivity introduces the difficulty that causal effects of non-focal individuals' behaviour depend on a focal individual's phenotype. There are at least three approaches to dealing with this challenge:

(i) One approach to $IF_{\text{Hamilton}}$, which we used, is to apply the original stripping procedure. That is, if a non-focal individual performs an act that causes the focal organism's reproduction to change (compared to if the act did not occur), then we calculate $IF_{\text{Hamilton}}$ as if this act did not occur (i.e. 'stripping'). This leads to the conclusion that $IF_{\text{Hamilton}}$ fails as a phenotypic maximand, because it unduly neglects a component of reproductive success that the focal individual *can* influence. Creel's paradox [12] neatly exemplifies the problem this creates when trying to account for obviously adaptive traits: $IF_{\text{Hamilton}}$ implies that it is better to be a helper than a breeder in a cooperative breeding system (Figure 3; Supplementary Material 3).

(ii) Alternatively, anticipating the inadequacy of approach (i) to capture all of a focal organism's causal effects, one might conclude that $IF_{\text{Hamilton}}$ simply cannot be applied in non-additive situations. This might be called the 'Grafen-Nowak approach', e.g. after refs [9] ("the question of how to define inclusive fitness in the absence of additivity has not been settled, and so fundamental theory on the non-additive case can hardly yet begin") and [29] ("since non-linear, synergistic phenomena cannot be attributed to individual actors, there is in general no meaningful way to define an individual's inclusive fitness").

(iii) One can abandon the task of calculating $IF_{\text{Hamilton}}$ as a property of an organism, and instead calculate the *inclusive fitness effect* of a focal gene or trait. This can be done with methods such as neighbour-modulated fitness (Supplementary Material 4) that automatically 'strip' appropriate components of only the effects of a particular gene or trait. One version of this approach, called the Taylor-Frank method [30,31], is very useful for constructing models, albeit without directly engaging with the phenotypic maximand concept. Another version, called the 'general form of Hamilton's rule' [32–34], defines a focal gene's inclusive fitness effect so as to make it positive by definition for any positively selected gene – even if it is a rogue gene that lowers vehicle quality. Although this formulation creates the impression of selection having a coherent direction, it does not resolve the question of how the opposing phenotypic effects of rogue genes and other genes play out in evolutionary time.

Approaches (i) and (ii) both support our conclusion that $IF_{\text{Hamilton}}$ is not a general phenotypic maximand; and approach (i) makes it explicit why $IF_{\text{Hamilton}}$ fails. Approach (iii) is silent on what, if any, property of an organism qualifies as a phenotypic maximand, as it is unconcerned with calculating *IF* as a property of an organism. Unfortunately, this limitation is frequently obscured by the practice of equating the *inclusive fitness effect* (applicable to a gene or trait) with inclusive fitness itself.

For example, consider a focal organism that produces *X* offspring, and causes its relatives of relatedness *r* to produce another *Y* offspring, by expressing several different traits. $IF_{\text{folk}}$ is readily defined as $X + rY$. But what is the focal organism's $IF_{\text{Hamilton}}$? According to approach (i), we can answer this question by measuring the component to be stripped, as the change in the focal organism's reproduction that would ensue from preventing all social acts of non-focal individuals. According to approach (ii), the question is meaningless unless all fitness interactions are additive, because the focal individual's $IF_{\text{Hamilton}}$ is not defined in the general case. And according to approach (iii), we cannot answer the question as the components to be stripped will differ from trait to trait, yielding no overall measure of $IF_{\text{Hamilton}}$ as a property of an individual.

Although $IF_{\text{Hamilton}}$ is the orthodox way to define inclusive fitness, we conclude that it is only a phenotypic maximand when interactions are additive. It only applies when the number of offspring which the social environment causes an individual to produce is unaffected by any aspect of the focal individual's phenotype that could be selected for [9]. In that special case, it makes no difference whether we think of $IF_{\mathrm{Hamilton}}$ or $IF_{\mathrm{folk}}$ as being maximised: they are both maximised by the same strategy, a point which has been made in a more general form by Okasha & Martens [25].

## 8. Discussion

The most profound achievement of evolutionary theory is to explain the origin of complex organismal design that was once attributed to supernatural creation. According to the theory of natural selection, complex design arises gradually because changes in numerous phenotypic dimensions, induced by many genes, are predominantly guided in a coherent direction. The guiding principle that gives directionality to this process was identified by Darwin [2] as "the improvement of each organic being in relation to its organic and inorganic conditions of life", and refined by Hamilton [1,6] as the improvement of inclusive fitness. Here we have tried to emphasise and strengthen these core ideas by modifying some of the theory's details.

One of these modifications bears on the fiery debate between critics and defenders of inclusive fitness ignited by Nowak et al. [35]. As we see it, both sides of the controversy make some valid claims. The critics are correct that inclusive fitness, when defined as $IF_{\mathrm{Hamilton}}$, is a meaningful property of individual organisms (and hence a candidate phenotypic maximand) only under narrow conditions. But the defenders of inclusive fitness are equally correct to counter that organismal design can be understood, under very general conditions, in terms of inclusive fitness maximisation [36]. We suggest that the discrepancy between these statements is resolved by replacing $IF_{\mathrm{Hamilton}}$ with $IF_{\mathrm{folk}}$, which, we have argued, is a more general maximand.

We advocate the idea that long-term phenotypic evolution tends to follow the genome's 'majority interest'. Our rationale is that, although only genes that actually affect a given trait matter for its evolution, the genes that matter can change over time [28,37]. Even if a trait is currently affected by only one or a few loci, in the long term the whole genome is a target for mutations whose effects can modify those of these few loci. This makes it relevant to ask what modifier genes would be selected for. Are they those that strengthen or those that undermine a given phenotypic effect? The phrase "the trait serves/opposes the genome's majority interest" is shorthand for: the trait selects for unlinked modifiers improving/undermining it. Accordingly, the genome's 'majority interest' (formally encapsulated in a reference gene's interest) should manifest over evolutionary time because traits that align with it tend to be improved through complementary, cumulative contributions from unlinked genes, whereas traits opposed to it will tend to be eliminated.

Fortunately, the invaluable 'Taylor-Frank method' [30] to construct kin selection models is fully compatible with our theory. This method finds evolutionarily stable values of a continuous trait such that no mutant gene can invade that slightly changes the resident trait value. This includes stability against small-effect, low-penetrance genes that meet our definition of a reference gene. Since only a population whose members already maximise $IF_{\mathrm{folk}}$ leaves no scope for the invasion of a reference gene (section 6), this implies – perhaps surprisingly – that the Taylor-Frank method finds strategies that (locally) maximise $IF_{\mathrm{folk}}$ rather than $IF_{\mathrm{Hamilton}}$. How could this important implication have been overlooked? We see two likely reasons. First, $IF_{\mathrm{Hamilton}}$ works as an accounting tool for genes with small (hence approximately additive) effects, which are the type of genes considered by the Taylor-Frank method. Some might therefore be tempted to conclude that $IF_{\mathrm{Hamilton}}$ will also work as a phenotypic maximand. This conclusion is unjustified, however, because (approximate) additivity at the gene level does not imply additivity at the organism level. And without additivity at the organism level, $IF_{\mathrm{Hamilton}}$ does not fully capture an organism's causal effects on gene propagation (section 7). Second, ambiguity arises from the widespread use of verbal definitions that purport to describe $IF_{\mathrm{Hamilton}}$, but, in fact, obfuscate $IF_{\mathrm{Hamilton}}$ and $IF_{\mathrm{folk}}$. For example, inclusive fitness has been called "the property of an individual organism which will appear to be maximized when what is really being maximized is gene survival" [38] or "the component of reproductive success an organism can influence" [39]. While the latter definition maps to $IF_{\text{Hamilton}}$ when applied to models with additive interactions, it maps to $IF_{\text{folk}}$ when applied to nature. In reality, no component of an organism's reproduction is *a priori* beyond its influence, in the sense that it is unaffected by any evolvable aspect of the focal organism's phenotype. For example, for an organism to convert help from others into offspring, it must allocate resources to its gonads. In nature, the "component of reproductive success an organism can influence" therefore includes all its offspring. Similarly, when applied to nature, Hamilton's [1] "generalized unrigorous statement of the main principle" (which does not mention any 'stripping') can arguably be read as an implicit endorsement of $IF_{\text{folk}}$: "The social behaviour of a species evolves in such a way that in each distinct behaviour-evoking situation the individual will seem to value his neighbours' fitness against his own according to the coefficients of relationship appropriate to that situation."

Our approach is inspired by the 'gene's eye view' of adaptation made popular by Dawkins' *The Selfish Gene* [4]. According to this view, adaptive phenotypes can be identified by metaphorically adopting a gene's first-person perspective to ask: "what phenotype should I induce to make more copies of myself?" However, in the words of Hammerstein [28] "…a naive interpretation of the idea of the 'selfish gene' can easily direct our attention to an inappropriate level of biological organization (genes instead of phenotypes). This is so because [in the multi-locus case] the genetic scene can only be described as an 'incredible mess' although very clear economic principles hold - in the long run - at the phenotypic level." Our reference gene concept is an attempt to tidy up the 'gene's eye view', by envisaging a gene that embodies the guiding principle of multi-locus evolution. This approach reflects the view of many biologists that it is usually more interesting to ask "what phenotypes are adaptive?" than to ask "what hypothetical gene could be selected for?" For example, undue focus on the latter question might lead us to predict fathers who kill their daughters to feed their sons (if caused by a gene on the father's Y- chromosome - a hypothetical variant of a *Mendelian outlaw gene* [20]); or to predict indiscriminate altruism between all members of a species (caused by a *greenbeard gene* gone to fixation). Such outcomes involving rogue genes are unlikely to be observed in nature because – being incompatible with the genome's majority interest – they can neither evolve through cumulative contributions of unlinked genes, nor be stable in the long term. We are left with two equivalent metaphors for long-term phenotypic evolution. We can think either of reference genes strategically 'trying' to maximise their propagation, or of organisms evolving to maximise their vehicle quality (or $IF_{\text{folk}}$). Both metaphors capture the view that organisms are integrated systems shaped over generations by the contributions of numerous genes, and, as such, are unlikely to perpetually retain traits under counter-selection from the majority of the genome.

To conclude, our present theory might confirm what many readers intuitively think – that organisms appear to be designed to maximise the weighted offspring count that defines $IF_{\text{folk}}$. The prevalence of this intuition is seen in the persistent tendency to define inclusive fitness as $IF_{\text{folk}}$ in teaching materials and other non-mathematical texts [16,40–42]. This view has, however, never been explicitly justified, and it stands in contradiction to the prevailing orthodoxy among theoreticians. Our line of argument, if valid, would create the unusual situation that orthodoxy should change to match the textbooks, rather than the other way around.

**Acknowledgements:** We thank David Queller (who went way beyond the call of duty as a referee), Jono Henshaw, Jussi Lehtonen, Piret Avila, and Jaakko Toivonen for discussions and comments on the manuscript; Tom Wenseleers, Zoltan Barta, Jutta Schneider, Mikael Puurtinen, Jannis Liedtke and Sara Calhim for comments on the manuscript; and Erol Akcay and Jeremy van Cleve for helpful criticism. **Funding:** Academy of Finland (LF; grant 283486) and the Australian Research Council (MDJ). **Author contributions:** LF had the idea and wrote the first draft. MDJ contributed through discussion of ideas and writing. **Competing interests:** We declare no competing interests. **Data and materials availability:** Code is available at doi:10.5061/dryad.h1c0b52.

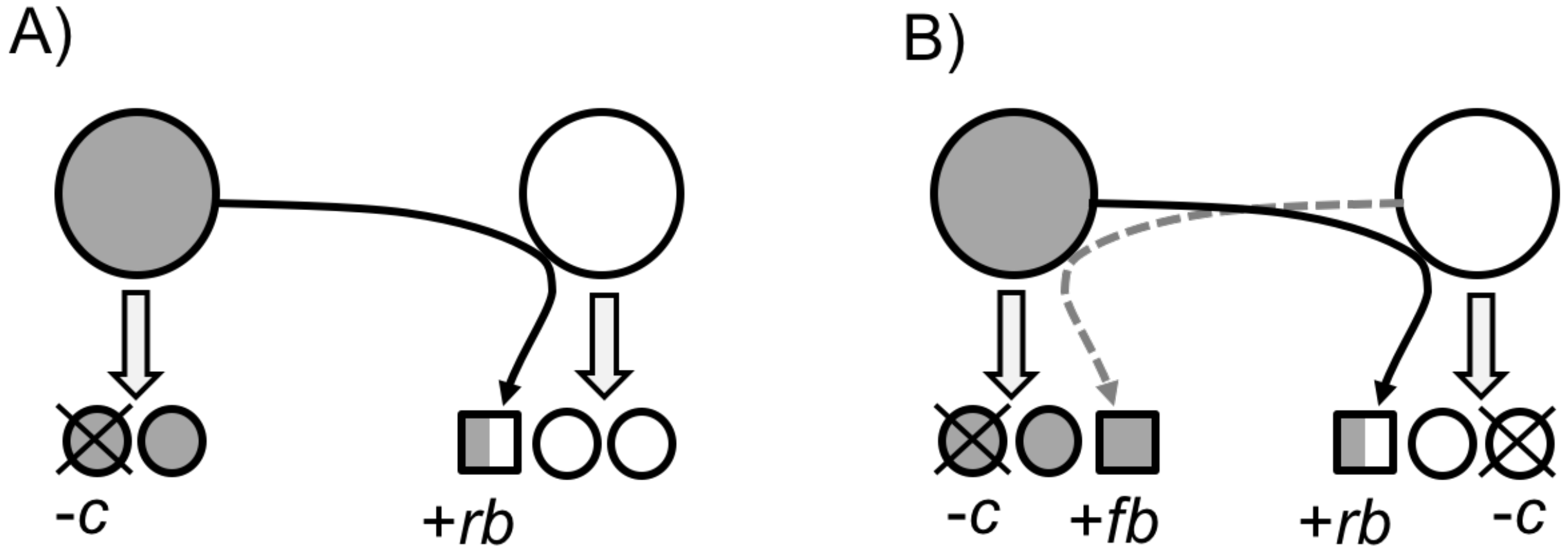

Figure 1. Performance of inclusive fitness measures as an accounting tool for genes without or with mirror effect. Big circles represent adults; the shaded one is the focal actor. Small circles represent offspring produced without help from the social environment. Crossed-out small circles represent offspring not produced as a result of a costly helping act. Small squares represent offspring produced through helping. The shading of small squares represents such offsprings' relatedness to the focal individual, relative to its own offspring. Black arrows represent helping acts performed by the focal individual, pointing to the resultant offspring produced by the non-focal individual. Dashed arrows represent helping acts received by the focal individual from its social environment. We compare $IF_{\text{Hamilton}}$ with $IF_{\text{folk}}$, which differs from $IF_{\text{Hamilton}}$ in that none of the focal individual's offspring are stripped away. A) In a population where by default each individual produces two offspring without giving or receiving help (*baseline* = 2), a mutant gene *without mirror effect* causes the focal individual to help a relative, yielding an indirect fitness benefit *rb*, at cost –*c*. Because the focal individual's behaviour is not mirrored by its relative, we have $IF_{\text{Hamilton}} = IF_{\text{folk}} = baseline + rb - c$, and the gene is positively selected if $IF > baseline$ (i.e. *rb* - *c* > 0). Thus, both $IF_{\text{Hamilton}}$ and $IF_{\text{folk}}$ work as an accounting tool for this type of gene. B) Similar to A, but *with mirror effect*: here the mutant gene which causes the focal individual to help is also expressed in any relatives that carry its identical copies. As a result, the focal individual produces $fb$ additional offspring, where $f = f[r, p]$ is the probability of receiving help, which is a function of relatedness *r*, the gene's frequency *p,* as well as the gene's penetrance. (Moreover, looking beyond the simplistic case where all helping in the population is due to the focal allele, the $f$ term should also account for help received due to behaviour encoded by other loci.) This situation yields $IF_{\text{Hamilton}} = baseline + rb - c$ (not including the $fb$ offspring produced due to the social environment) and $IF_{\text{folk}} = baseline + rb - c + fb$. Now $IF_{\text{Hamilton}} > baseline$ still correctly predicts selection on the focal gene (provided fitness effects are additive [26]), because it isolates the gene's causal effects from the correlational component $fb$ that would arise even if the gene in the focal organism were not expressed. By contrast, $IF_{\text{folk}} > baseline$ does *not* correctly predict selection because the term $fb$ includes a benefit (in the focal individual) whose cost (in another individual) is unaccounted for [16]. However, rather than being a shortcoming of $IF_{\text{folk}}$, this merely reflects the general difficulty of inferring a causal effect from correlational data. In a causal modelling framework [17,18], this difficulty is readily avoided by calculating $\widehat{IF}_{\text{folk}}[\text{don't help}] = baseline + fb$ as the focal individual's counterfactual value of $IF_{\text{folk}}$ that would arise if the focal individual did not help. Then the focal gene's causal effect on the focal organism's $IF_{\text{folk}}$ is positive if $IF_{\text{folk}}[\text{help}] - \widehat{IF}_{\text{folk}}[\text{don't help}] = rb - c > 0$, which recovers the correct inclusive fitness effect. Thus, the focal gene is selected for if expressing it causes the focal organism's $IF_{\text{folk}}$ to increase.

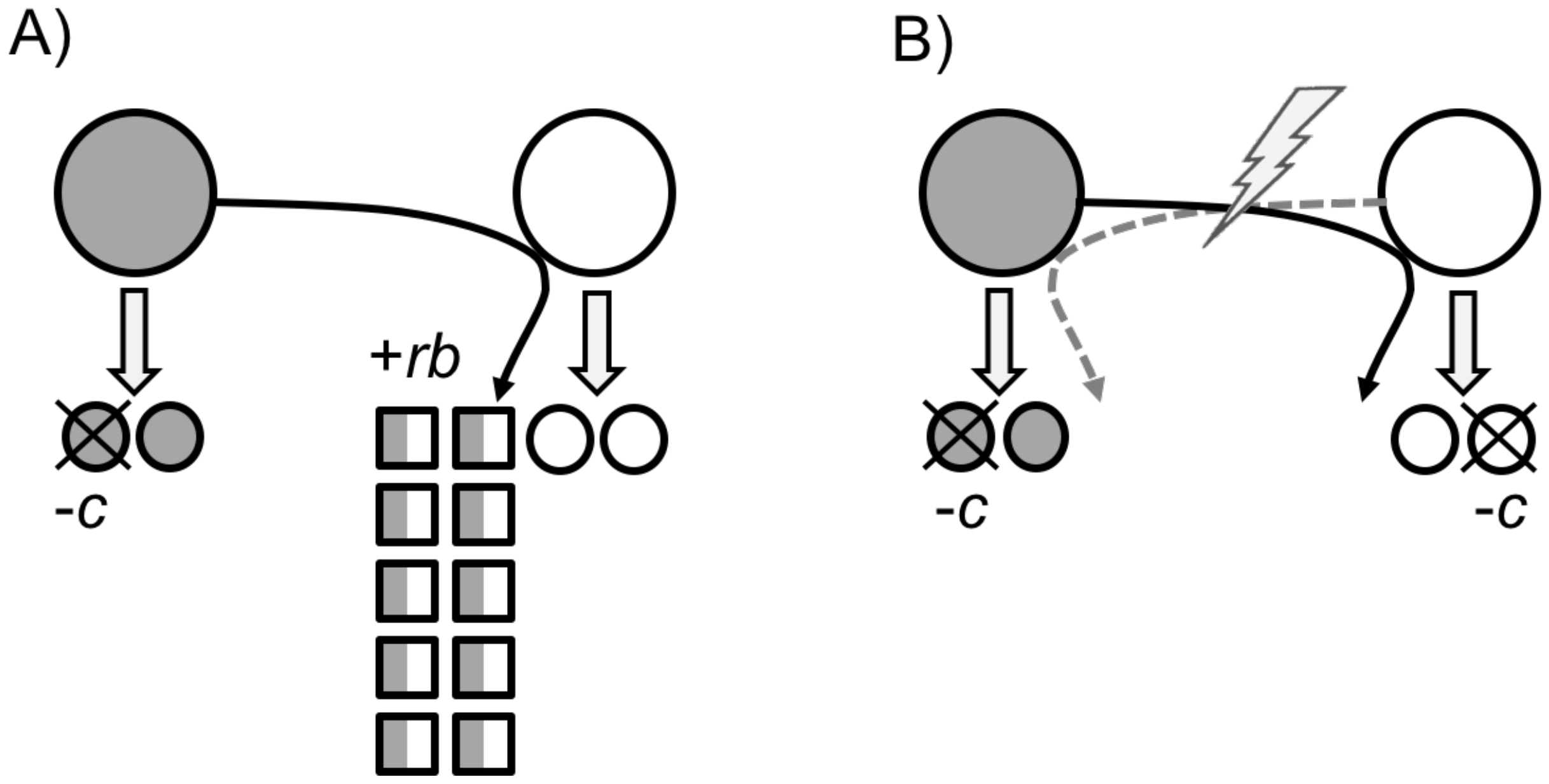

Figure 2. Example of how the mirror effect, in combination with non-additive interactions between individuals, can generate selection for a trait that reduces vehicle quality. Consider a population where siblings interact (e.g., pedigree relatedness $r = 0.5$), and where unilateral help (A) is highly efficient (e.g., $b = 10$, $c = 1$) whereas mutual help (B) is completely inefficient due to strong interference between matching phenotypes (symbolised by lightning bolt; $d = -10$ in the notation of Supplementary Material 1). In this situation, helping cannot evolve based on a gene with full penetrance, because benefits accrue exclusively to individuals who lack the helping gene. Thus, when a full-penetrance helping gene (which is subject to the mirror effect) is introduced at low frequency into the population, its alternative allele (which can be considered a full-penetrance non-helping gene) will quickly spread back to fixation. This occurs even though at the phenotypic level, individuals could increase their vehicle quality by switching to unilateral helping, thus reaping the indirect benefits shown in A. Even though defection to non-helping reduces vehicle quality, it spreads to fixation based on a *mirror effect rogue gene* - leading to an equilibrium where helping does not occur. In other words, organisms end up making no use of the huge indirect fitness benefit that would accrue from unilateral helping, which contradicts the idea that individuals are selected to maximise their $IF_{\text{folk}}$. Crucially, however, this equilibrium without helping is only stable under the restrictive assumption that mutations without mirror effect cannot arise (Supplementary Material 1). If such mutants arise (e.g. a low-penetrance gene; or a gene for helping your younger sibling, conditional on being the older one), they generate selection for helping due to the indirect benefits shown in A.

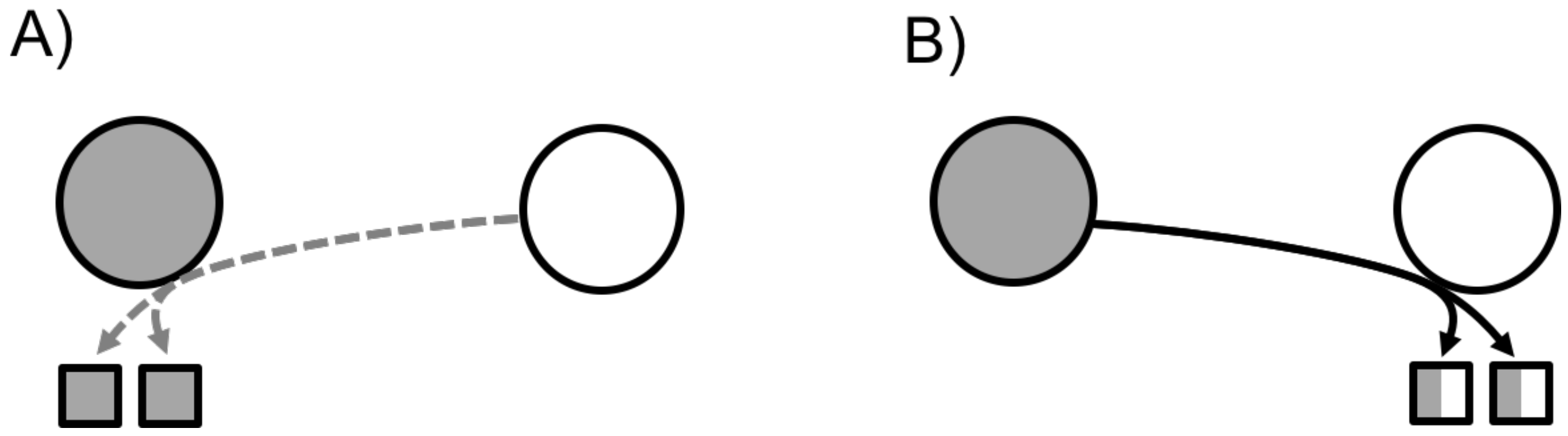

Figure 3. Creel's paradox, modified after Queller [12]: in an obligate cooperative breeding system where reproduction requires exactly one breeder and one helper, the focal individual has a choice between becoming the breeder (A) or the helper (B), while the non-focal individual (based on some asymmetry) must take the remaining role. Since offspring produced due to the social environment are excluded from $IF_{\mathrm{Hamilton}}$, the focal individual has lower $IF_{\mathrm{Hamilton}}$ in A than B (0 versus $2r$), despite transmitting more genes as a breeder. Invoking $IF_{\mathrm{Hamilton}}$ as a phenotypic maximand predicts wrongly that the focal individual should prefer to become the helper (Supplementary Material 3). By contrast, the focal individual's $IF_{\mathrm{folk}}$ is higher in A than B (2 versus $2r$), predicting correctly that the focal individual should prefer to become the breeder. This matches Queller's [12] prediction, which he obtained (without invoking inclusive fitness as a property of an organism) by applying Hamilton's rule separately to two genes, each expressed conditionally in one of the two roles.

**Supplementary Material 1: Mirror effect rogue genes**

We have defined mirror effect rogue (MER) genes as alleles which reduce the vehicle quality of the individuals expressing them, but spread because of the mirror effect. Here we use a model to examine under what circumstances MER genes can exist, and to what extent they may then pose an impediment to $IF_{\text{folk}}$ being a phenotypic maximand. First, we characterise the conditions that select for behaviours encoded by genes with mirror effect. Next, we characterise the conditions that select for behaviours encoded by genes without mirror effect (e.g., a reference gene that is guaranteed to be positively selected if it increases vehicle quality, since vehicle quality is defined by a reference gene's propagation success). Then, we compare the two sets of conditions to identify conditions where selection on genes with or without mirror effect favours opposite phenotypes. These are the conditions where MER genes can occur. Finally, we show that equilibria established by MER genes (at which $IF_{\text{folk}}$ is not maximised) are not stable against invasion by mutant genes without mirror effect. In contrast, the corresponding equilibria at which $IF_{\text{folk}}$ is maximised are stable against mutant genes both with and without mirror effect.

Consider a haploid species (for simplicity) with the following life cycle: individuals interact for one round of a pairwise game, played between relatives of pedigree relatedness $r$. For example, everyone interacts once with a full sibling ($r = 0.5$), or everyone interacts once with a half-sibling ($r = 0.25$), or everyone interacts once with either a clone (with probability $r$) or with an unrelated individual. The essential point is that a rare gene, if present in a focal individual, occurs in its social partner with probability $r$ due to coancestry. After this pairwise interaction, individuals disperse randomly, mate, and reproduce. The assumption of random dispersal rules out local competition (see Supplementary Material 5, Q34), such that all offspring have an equal chance to reproduce. This ensures that offspring number is an evolutionarily relevant measure of reproductive success. There are two behavioural options: cooperate (denoted "$+$") or defect (denoted "$-$"). If a focal individual cooperates, it pays cost $c$ to provide to its relative either benefit $b$ (if the relative defects) or $b + d$ (if the relative cooperates). If the focal individual defects, it pays no cost nor does it provide a benefit. If there is synergy ($d > 0$), mutual cooperation is more efficient than unilateral cooperation. If there is interference ($d < 0$), unilateral cooperation is more efficient than mutual cooperation. If fitness effects are additive ($d = 0$), mutual and unilateral cooperation are equally efficient. Although here we focus on the evolution of a cooperative trait (with $b > 0$, $c > 0$), an analogous argument holds for a selfish trait (with $b < 0$, $c < 0$).

| | | non-focal actor $+$ | non-focal actor $-$ |
|---|---|---|---|
| focal actor | $+$ | $b + d - c$ | $-c$ |
| focal actor | $-$ | $b$ | $0$ |

The focal individual's resultant payoffs, as listed in the matrix above, are changes in direct reproductive success: i.e., own offspring produced (or not produced) as a result of the interaction, as compared to some baseline number.

*1. Genes with mirror effect*

Consider a gene which always causes its carriers to cooperate, whereas its allele always causes its carriers to defect. This type of gene is subject to a mirror effect of maximum strength: if two individuals that have the focal gene interact, both are certain to express the gene (hence to cooperate). In a population where relatives interact, this type of gene makes its carriers interact disproportionally with their own type. Specifically, let relatedness *r* cause phenotypic correlation $R = r$ between social partners, such that the probability of facing a given phenotype is conditional on one's own phenotype as follows [43]: cooperators face a cooperator with probability $f_+ = R + (1 - R)p$, while facing a defector with probability $(1 - f_+)$. Here, $R$ is the probability that a non-focal individual 'mirrors' a focal individual's phenotype because their genes at the focal locus are identical by descent; and $p$, the frequency of cooperation in the population, corresponds to the probability that a focal cooperator faces a cooperator even when their genes at the focal locus are not identical by descent. Defectors face a cooperator with probability $f_- = (1 - R)p$, while facing a defector with probability $= (1 - f_-)$. This leads to expected payoffs $W_+ = f_+(b + d - c) + (1 - f_+)(-c)$ for cooperators and $W_- = f_-(b)$ for defectors. Since we are dealing with genes that are always expressed, personal payoffs of each phenotype are representative of the underlying genes' transmission success. Hence we can use personal payoffs of each phenotype to infer the direction of selection. If $W_+ = W_-$, the focal gene for cooperation is selectively neutral. Solving for $p$, this occurs at equilibrium frequency

$$\hat{p} = \frac{c - r(b+d)}{d(1-r)} \qquad (1).$$

Likewise, the focal gene for cooperation is selected positively when $W_+ > W_-$, and negatively when $W_+ < W_-$. By substituting into these inequalities, we can characterise selection as follows. Given synergy ($d > 0$), cooperation is selected positively while $p > \hat{p}$ and negatively while $p < \hat{p}$. This implies that, if $\hat{p}$ is an internal equilibrium (i.e., in the range $0 < \hat{p} < 1$), it is unstable due to positive frequency-dependent selection. Given interference ($d < 0$), cooperation is selected positively while $p < \hat{p}$ and negatively while $p > \hat{p}$. This implies that, if $\hat{p}$ is an internal equilibrium, it is stable due to negative frequency-dependent selection. This is the equilibrium found in earlier studies [24,25] where $IF_{\text{folk}}$ was *not* maximised. Following Hines & Maynard Smith [44], we call this the Grafen ESS.

In a population at the Grafen ESS, consider a rare mutant gene (either an allele at the same locus, or a mutation at a second locus overriding the first) that makes its carriers cooperate with probability *P*, and defect otherwise. If we think of defection as the 'null' phenotype, we can interpret this mutant gene as a cooperation gene with penetrance *P*. When a mutant individual who carries this gene faces a cooperator (as happens with probability $f_{\text{mut}} = rP + (1 - r)\hat{p}$ ), it obtains payoff $P(b + d - c) + (1 - P)b$. When facing a defector (with probability $(1 - f_{\text{mut}})$), the mutant obtains $P(-c)$. Hence the mutant's expected payoff is $W_{\text{mut}} = f_{\text{mut}}[P(b + d - c) + (1 - P)b] + (1 - f_{\text{mut}})P(-c)$. The mutant gene can invade if mutant individuals enjoy higher neighbour-modulated fitness than the population mean, i.e. if $W_{\text{mut}} > W_+ = W_-$. For *P* in the range ($0 < P < 1$), this reduces to $d < 0$, which is always met when the Grafen ESS exists. Thus, any cooperation gene with incomplete penetrance can invade the Grafen ESS.

*2. Genes without mirror effect*

In a population where phenotypes are controlled by a one-locus two-allele system with full penetrance, such that interacting phenotypes tend to resemble each other (see above), consider selection at a second locus unlinked to the first. At this second (previously neutral) locus, a rare mutant gene variant arises that encodes defection without mirror effect, which transforms individuals that would otherwise have cooperated into defectors. There are two ways in which such a mutant gene may arise: (i): a neutral gene mutates into an allele for defection with low penetrance (thus meeting our definition of a reference gene); (ii): a neutral gene mutates into an allele for defection whose expression is conditional on some asymmetry between organisms.

Consider a focal individual that is turned into a defector due to expressing a rare mutant gene without mirror effect. When this focal individual faces a cooperator (as happens with probability $f_+$, since the focal individual would have cooperated but for the focal gene's effect), it obtains payoff $b$ instead of $b + d - c$, amounting to net gain $-d + c$. If the focal individual faces a defector (with probability $1 - f_+$), it obtains payoff 0 instead of $-c$, amounting to net gain $c$. Thus, the focal gene's causal effect on the focal individual's change in direct fitness, as a result of defecting, is $\Delta_{\text{direct}} = f_+(-d + c) + (1 - f_+)c$. Since a gene without mirror effect is not expressed by every individual that carries it, its transmission success cannot be inferred from the personal payoff of only the individuals that do express it. Instead, the payoffs of individuals that carry the focal gene but do not express it must also be accounted for. We do this by considering indirect effects, mediated by the relative's payoff: if the relative is a cooperator, it obtains (due to the focal individual's change in behaviour) $-c$ instead of $b + d - c$, amounting to net change $-b - d$. If the relative is a defector, it obtains 0 instead of $b$, amounting to net change $-b$. Thus, the focal gene's causal effect on the focal individual's change in indirect fitness, as a result of defecting, is $\Delta_{\text{indirect}} = r[f_+(-b - d) + (1 - f_+)(-b)]$. The focal gene encoding defection is selectively neutral if it has zero net effect on the number of copies transmitted to the next generation. This net effect includes all causal effects of the focal gene being expressed (compared to the counterfactual of not being expressed) in the focal individual. Selective neutrality occurs when $\Delta_{\text{direct}} + \Delta_{\text{indirect}} = 0$; i.e., when expressing the focal gene does not change the focal individual's vehicle quality. This occurs when $p$ equals

$$p^- = \frac{c - r(b + d + rd)}{d(1 - r^2)} \qquad (2).$$

Likewise, when $\Delta_{\text{direct}} + \Delta_{\text{indirect}} > 0$, the focal gene is selected positively because it causes more copies to be transmitted to the next generation (compared to the number transmitted in the absence of its phenotypic effect; i.e. compared to a neutral gene). And when $\Delta_{\text{direct}} + \Delta_{\text{indirect}} < 0$, the focal gene is negatively selected for analogous reasons. By substituting into these inequalities, we can characterise selection as follows. Given synergy ($d > 0$), the focal gene is selected positively if $p < p^-$ (i.e., the frequency of cooperators is sufficiently low) and negatively if $p > p^-$. Given interference ($d < 0$), the focal gene is selected positively if $p > p^-$ (i.e., the frequency of cooperators is sufficiently high) and negatively if $p < p^-$.

Conversely, now consider selection for a rare gene encoding cooperation without mirror effect, which transforms individuals that would otherwise have defected into cooperators. If a focal individual expressing this gene faces a cooperator (as happens with probability $f_-$ , since the focal individual would have defected but for the focal gene's effect), it obtains payoff $b + d - c$ instead of $b$, amounting to net gain $d - c$. If the focal individual faces a defector (with probability $1 - f_-$), it obtains payoff $-c$ instead of 0, amounting to net gain $-c$. Thus, the focal gene's causal effect on the focal individual's change in direct fitness, as a result of defecting, is $\Delta_{\text{direct}} = f_-(d - c) + (1 - f_-)(-c)$. Now consider indirect effects, mediated by the relative's payoff: if the relative is a cooperator, it obtains (due to the focal individual's change in behaviour) $b + d - c$ instead of $-c$, amounting to net gain $b + d$. If the relative is a defector, it obtains $b$ instead of 0, amounting to net gain $b$. Thus, the focal gene's causal effect on the focal individual's change in indirect fitness, as a result of cooperating, is $\Delta_{\text{indirect}} = r[f_-(b + d) + (1 - f_-)(b)]$. The focal gene encoding cooperation is selectively neutral if has zero net effect on its number of copies transmitted to the next generation. This net effect includes all causal effects of the focal gene being expressed (compared to the counterfactual of not being expressed) in the focal individual. Selective neutrality occurs when $\Delta_{\text{direct}} + \Delta_{\text{indirect}} = 0$. This occurs when $p$ equals

$$p^+ = \frac{c - rb}{d(1 - r^2)} \qquad (3)$$

Using the same logic as above, we can characterise selection as follows. Given synergy ($d > 0$), the focal gene is selected positively if $p > p^+$ (i.e., the frequency of cooperators is sufficiently high) and negatively if $p < p^+$. Given interference ($d < 0$), the focal gene is selected positively if $p < p^+$ (i.e., the frequency of cooperators is sufficiently low) and negatively if $p > p^+$. An alternative method to obtain eqns. (2) and (3) is to calculate selection for a modifier gene with general penetrance level $P$, and then take the limit of low penetrance ($P \to 0$).

*3. Mirror effect rogue genes*

By our definition, a MER gene spreads via the mirror effect despite (on average) reducing the vehicle quality of the individuals expressing it. This occurs when a trait is positively selected as described in 'Genes with mirror effect' above, while the opposite trait is positively selected as described in 'Genes without mirror effect' (e.g. based on a reference gene). The conditions for this to occur simultaneously can only be met when there is interference, $d < 0$ (Table S1). Specifically, a MER gene for defection can spread at some $p > \hat{p}$ whenever pure cooperation is not an ESS (e.g. Figure 2). Likewise, a MER gene for cooperation can spread at some $p < \hat{p}$ whenever pure defection is not an ESS. Intuitively, these findings can be explained as follows. Interference reduces the efficiency of cooperating with other cooperators. This creates conditions where switching to cooperation is worthwhile only if it can be done unilaterally, but not if it involves a correlated switch (due to the mirror effect) by the social partner (Figure 2). Likewise, there are conditions where switching to defection is worthwhile only if it can be done unilaterally, but not if the switch is mirrored by relatives.

It is, perhaps, not obvious why MER genes do not occur under synergy ($d > 0$), even though the mirror effect broadens the conditions under which a gene for cooperation can invade (from $rb - c > 0$ without mirror effect, to $rb - c + rd > 0$ with mirror effect [25]). In the range where only the latter condition holds, the mirror effect evidently reverses the direction of selection, but it does so *without* reducing cooperators' vehicle quality. Intuitively this can be explained as follows: the mirror effect elevates the (local) frequency of cooperators around any focal cooperator, up to the point where cooperation becomes optimal given positive frequency-dependent selection.

*4. Reciprocal invasion of genes with or without mirror effect*

What phenotypes will evolve in the long run, if predicted equilibria differ based on genes with or without mirror effect? This depends, in part, on whether each equilibrium can be invaded by genes of the other type. In what follows we assume $d < 0$, as required for stable mixed equilibria to exist. Because a gene for cooperation without mirror effect is selected for if $p < p^+$ (see above), the Grafen ESS at $\hat{p}$ (with mirror effect) can be invaded by a gene for cooperation without mirror effect if $\hat{p} < p^+$. This yields $rb - c + d(1 + r) < 0$, which is the condition for pure cooperation not being an ESS [25]. Thus, whenever pure cooperation is not an ESS, the Grafen ESS can be invaded by a gene for cooperation without mirror effect. Similarly, the Grafen ESS can be invaded by a gene for defection without mirror effect if $\hat{p} > p^-$. This yields $rb - c > 0$, which is always true when $\hat{p} > 0$ in the first place. Thus, the Grafen ESS can always be invaded by a gene for defection without mirror effect.

Under the same parameter settings, two kinds of equilibrium – symmetric and asymmetric – can exist at which vehicle quality is maximised, such that mutant genes without mirror effect cannot invade. Here we do not model explicitly how these equilibria might be reached (but see Supplementary Material 2). Instead, we merely note that eventually one of them should be reached if phenotypic evolution follows the genome's 'majority interest' towards phenotypically adaptive outcomes. Consistent with the 'streetcar theory of evolution' [28], we show in Supplementary Material 2 that, barring genetic constraints, phenotypically adaptive outcomes can become realised through a variety of genetic mechanisms.

*4.1 Symmetric ESS*

If there exists no asymmetry (or negotiation) between interacting individuals that would allow for conditional gene expression, then the mirror effect can nevertheless be avoided by genes having low penetrance. Successive invasions of such genes will tend to reduce the phenotypic correlation towards $R = 0$, so that $f_+ = f_- = p$ (i.e., the probability of facing a cooperator is independent of the focal individual's phenotype, and equals the frequency of cooperators in the population). Re-calculating either $p^+$ or $p^-$ given $R = 0$ yields the mixed ESS

$$p^* = \frac{c-rb}{d(1+r)} \qquad (4)$$

as the value of $p$ at which further mutants without mirror effect cannot obtain a selective advantage by switching phenotypes one way or the other. We call this equilibrium the standard ESS, to distinguish it from the Grafen ESS. The standard ESS may be approached in phenotypic space by the combined action of genes with and without mirror effect, where genes with successively weaker mirror effect do the 'fine-tuning' near the equilibrium (Supplementary Material 2). Alternatively, in what Grafen [24] called the 'continuous strategy case', the standard ESS can also be reached if evolution proceeds exclusively by small-effect genes affecting the propensity to cooperate. In a population at the standard ESS, the average payoff is $\overline{W} = p^*W_+ + (1-p^*)W_-$, where $W_+ = p^*(b+d-c) + (1-p^*)(-c)$ and $W_- = p^*(b)$ are the payoffs of cooperators and defectors, respectively. In this population, a mutant individual carrying a full-penetrance gene for cooperation (i.e. with mirror effect) obtains payoff $\widehat{W}_+ = \hat{f}_+(b+d-c) + (1-\hat{f}_+)(-c)$, where $\hat{f}_+ = r + (1-r)p^*$. The resident population is stable against this mutant if $\overline{W} > \widehat{W}_+$, which leads to

$$rb - c + d(1+r) < 0 \qquad (5).$$

This is the condition for pure cooperation not being an ESS, which is always satisfied when the standard ESS exists. Similarly, a mutant individual carrying a full-penetrance gene for defection obtains payoff $\widehat{W}_- = \hat{f}_-(b)$, where $\hat{f}_- = (1-r)p^*$. The resident population is stable against this mutant if $\overline{W} > \widehat{W}_-$, which leads to

$$rb - c > 0 \qquad (6).$$

This is the condition for pure defection not being an ESS, which is always satisfied when the standard ESS exists [25].

*4.2 Asymmetric ESS*

Alternatively, the mirror effect can be avoided by genes being expressed conditional on some (perhaps arbitrary) asymmetry between individuals. In the present model, an asymmetric ESS exists at which individuals cooperate in role A and defect in role B, such that $f_+ = 0$ and $f_- = 1$. For this outcome to be stable, two conditions need to be met. Firstly, it must be optimal to play "+" in role A given the individual in role B plays "–". This is the case when playing "+" instead of "–" in role A yields higher vehicle quality; i.e., $\Delta_{\text{direct}} + \Delta_{\text{indirect}} > 0$, where $\Delta_{\text{direct}} = -c$ and $\Delta_{\text{indirect}} = rb$. This recovers condition (6). Secondly, it must be optimal to play "–" in role B given the individual in role A plays "+". This is the case when playing "–" instead of "+" in role B yields higher vehicle quality; i.e., $\Delta_{\text{direct}} + \Delta_{\text{indirect}} > 0$, where $\Delta_{\text{direct}} = -d + c$ and $\Delta_{\text{indirect}} = r(-b-d)$. This recovers condition (5). The average payoff in a population using this asymmetric ESS is $\overline{W} =$ $pW_+ + (1-p)W_-$, where $p = 0.5$ (as implied by interactions occurring in pairs), and $W_+ = -c$; $W_- = b$ are the payoffs of cooperators and defectors, respectively. Now consider a mutant full-penetrance gene for cooperation (i.e. with mirror effect), whose carriers experience roles A or B with equal probability. When in role A, such a mutant behaves like an individual of the resident population (a 'resident'), but, unlike a resident, receives help with probability $r$. Its payoff in role A is thus $b + d - c$ with probability $r$, and $-c$ otherwise. When in role B, the mutant always faces a cooperator, yielding payoff $b + d - c$. The mutant's expected payoff is thus

$$\widehat{W}_+ = \frac{r(b+d-c)+(1-r)(-c)+b+d-c}{2}.$$

The resident population is stable against this mutant if $\overline{W} > \widehat{W}_+$. This recovers condition (6), which is one of the conditions for the asymmetric ESS to exist in the first place. Thus, the population is stable against cooperator mutations with mirror effect whenever it is stable against cooperator mutations without mirror effect.

Similarly, consider a mutant full-penetrance gene for defection. When in role A, a mutant individual carrying this gene always faces a defector, yielding the payoff from mutual defection, 0. When in role B, it faces a defector with probability $r$ (yielding payoff 0) and a cooperator with probability $(1-r)$ (yielding payoff $b$). The mutant's expected payoff is thus

$$\widehat{W}_- = \frac{(1-r)(b)}{2}.$$

The resident population is stable against this mutant if $\overline{W} > \widehat{W}_-$. This recovers condition (5), which is one of the conditions for the asymmetric ESS to exist in the first place. Thus, the population is stable against defector mutations with mirror effect whenever it is stable against defector mutations without mirror effect.

*5. Conclusion*

If social interactions are subject to interference between matching phenotypes (e.g., mutual help is less efficient than unilateral help; $d < 0$), MER genes may establish an evolutionary equilibrium at which individuals do not maximise vehicle quality. Crucially, however, this equilibrium can be invaded by genes without mirror effect (and indeed by genes with imperfect penetrance of any degree). In contrast, the reciprocal invasion by mutant genes with mirror effect, of the corresponding equilibria where vehicle quality is maximised, is not possible. Thus, only equilibria where vehicle quality ($IF_{\text{folk}}$) is maximised can exhibit phenotypic long-term stability [28] with respect to mutations with any level of penetrance.

Table S1: Conditions for the occurrence of MER genes

| MER gene for | $d$ | condition | comment |
|---|---|---|---|
| cooperation | $> 0$ | $p^- > p > \hat{p}$ | not satisfiable[1] |
| | $< 0$ | $p^- < p < \hat{p}$ | satisfiable if $rb - c > 0$, i.e. whenever pure defection is not an ESS [25] |
| defection | $> 0$ | $p^+ < p < \hat{p}$ | not satisfiable[2] |
| | $< 0$ | $p^+ > p > \hat{p}$ | satisfiable if $rb - c + d(1+r) < 0$, i.e. whenever pure cooperation is not an ESS [25] |

$p$ is the frequency of cooperators in the population; $\hat{p}$ is the frequency of cooperators at the Grafen ESS of eq. (1); $p^-$ (and $p^+$) are threshold values of $p$ at which a rare gene without mirror effect for turning cooperators into defectors (or defectors into cooperators) is selectively neutral.

[1] A contradiction arises because $p^- > \hat{p}$ implies $rb - c > 0$, while $p^- > 0$ implies $rb - c + rd(1 + r) < 0$. These cannot both be true if $d > 0$.

[2] A contradiction arises because $p^+ < \hat{p}$ implies $rb - c + d(1 + r) < 0$, while $p^+ < 1$ implies $rb - c + d(1 + r^2) > 0$. Since $r \leq 1$, these cannot both be true while $d > 0$. Note that $p^+ < 1$ is a necessary condition for $p^+ < p$ to hold, as $p$ cannot exceed 1.

**Supplementary Material 2: Individual-based simulations**

To illustrate the co-evolutionary interplay of genes with different levels of penetrance, we study the 'symmetric case' of Supplementary Material 1 with individual-based simulations. The simulation proceeds in $t_{max}$ discrete time steps.

*Genes and phenotypes.* We assume haploid genetics. Each individual has one main locus and $2*m$ modifier loci. The main locus has allelic values "0" for "defect" and "1" for "cooperate". Modifier loci have allelic values of "0" for "inactive" and "1" for "active". Half of the $2*m$ modifier loci are dedicated to modifying each of the two possible allelic values (0 or 1) of the main locus. In addition to specifying a phenotype (i.e., a 'default' phenotype to be expressed in the absence of modifiers), each allele at the main locus also has a property $M$ (called modifiability) that specifies its susceptibility to having its default phenotype changed by modifier genes. We consider scenarios where $M$ is either held fixed or free to evolve.

(a) Fixed modifiability. $M$ is constrained to always take the same value. In different simulations, this value can be either 0 ("no modification possible"; in effect, this simulates a one-locus system) or 1 ("maximum modifiability").

(b) Evolving modifiability. For each allele copy at the main locus, $M$ is initialised by sampling values from a continuous uniform distribution between 0 and 2. (The same distribution is also used to sample mutations – see below.) Although values $M > 1$ are functionally equivalent to $M = 1$ (see below), defining the range of $M$ in this way is useful for detecting selection for reduced $M$, as compared to the expected mean of $M = 1$ in the absence of selection.

If an individual's main allele has value "0" and modifiability $M$, the individual cooperates with probability $\min[M, 1] \cdot \sum v/m$, where $\sum v/m$ is the mean of the allelic values $v$ of the relevant modifier genes. Similarly, if an individual's main allele has value "1" and modifiability $M$, the individual cooperates with probability $1 - \min[M, 1] \cdot \sum v\ /m$. Whether the phenotype is to cooperate or defect is then randomly assigned based on these probabilities.

This formulation implies that a single active modifier gene on its own has probability $M/m$ of reversing the default phenotype specified by the main locus. Hence increasing $m$ is equivalent to decreasing the penetrance of each modifier gene.

*Mating and social interactions.* In each step, the total population of size $N$ is randomly arranged into $N/2$ mating pairs. Each pair sexually produces 4 offspring, that inherit alleles by unlinked Mendelian inheritance. Each set of offspring is arranged in 2 sibling-pairs, to play one round of the non-additive game of Supplementary Material 1. Payoffs, which define each individual's direct reproduction (i.e., neighbour-modulated fitness), are assigned based on interacting phenotypes.

*Mutation.* Mutations occur independently at each locus (of each offspring) with probability $\mu$. When a gene mutates, its allelic value switches from 0 to 1 or vice versa. In the continuous strategy case of Figure S1, new allelic values are sampled from a continuous uniform distribution between 0 and 1. In case (b), when $M$ is allowed to evolve freely, for each mutation at the main locus a new $M$ value is randomly sampled from a uniform distribution between 0 and 2. This formulation allows that high-penetrance mutations at the main locus, which override any modifiers, may arise at any time.

*Recruitment.* The next generation is obtained by randomly sampling $N$ offspring, using payoffs as sampling probabilities. To ensure that expected contributions to the future population (i.e. reproductive values) are proportional to relative payoffs, we sample *with* replacement. Rather than implying that the same individual can survive twice, this should be interpreted as shorthand for letting each individual reproduce many offspring in proportion to its payoff, and then randomly pick $N$ survivors from the resultant total pool.

*Default settings.* $t_{max}$=1000; $N$=1000; $\mu = 0.001$. Game payoff parameters: $b$ = 5; $d$ = -2.5; $c$ = 1. Relatedness is held fixed at $r$ = 0.5, as implied by interactions occurring between siblings. With these parameter settings, the Grafen ESS (Supplementary Material 1, eq. 1), at which expected $IF_{\text{folk}}$ is not maximised, occurs at cooperation frequency $\hat{p} = 0.2$. The corresponding standard ESS (Supplementary Material 1, eq. 4), at which $IF_{\text{folk}}$ is maximised, occurs at cooperation frequency $p^* = 0.4$. The population starts at cooperation frequency $p$ = 0.5.

*Results.* Figures show mean values of 100 replicate runs. Running the simulation as a one-locus two-allele system with full penetrance (i.e., $M$ = 0) leads to the Grafen ESS of eq. (1) (Fig. S1). Introducing mutations with incomplete penetrance quickly changes the outcome to the standard ESS of eq. (4) (Fig. S1, after time = 400). When running the simulation as a multi-locus system (i.e., $M \geq 0$), increasing the number (hence decreasing the penetrance) of modifiers shifts the equilibrium frequency of cooperation successively closer to the standard ESS, which is reached around $m$ = 5 when $M$ does not evolve [case (a)] (Figs. S2 A&B). In case (b), however, where the modifiability ($M$) at the main locus evolves, even a single modifier locus ($m$ = 1) for each of the two possible allelic values (0 or 1) of the main locus suffices to establish the standard ESS (Fig. S3). In this case $M$ evolves to lower values, which have the effect of limiting a modifier gene's penetrance. This is advantageous because it reduces the disadvantageous tendency of facing one's own phenotype (given interference, $d$ < 0) for individuals that carry a modifier gene. By contrast, when $M$ evolves and there are $m$ = 10 low-penetrance modifier genes, $M$ is selected to ensure full modifiability. This is because modifier genes automatically have low penetrance in this case, so being susceptible to them reduces the disadvantageous tendency of facing one's own phenotype. Consistent with the results of Supplementary Material 1, these results indicate that the Grafen ESS tends to be replaced by the standard ESS when the restrictive genetic assumptions of the 'discrete strategy case' are relaxed.

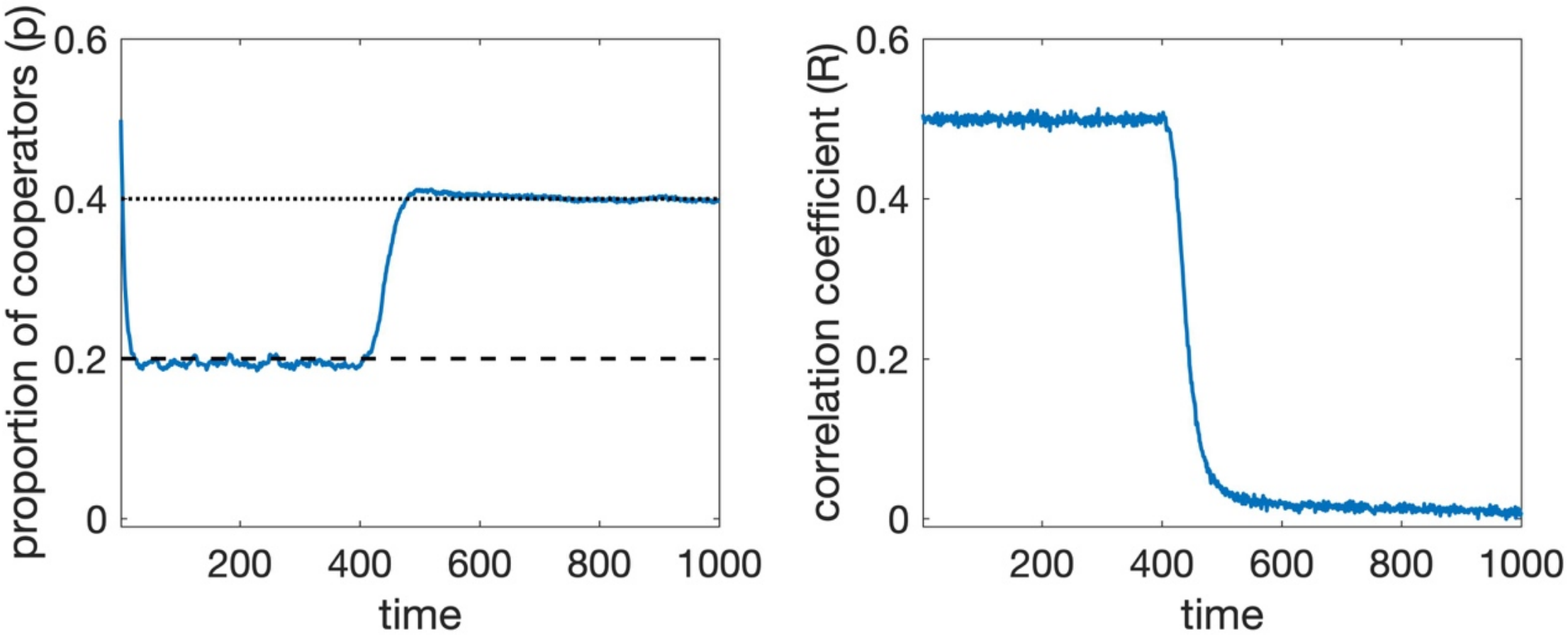

Figure S1. Discrete and continuous strategy case in a one-locus system. The main locus cannot be modified (i.e., $M$ = 0), so other loci have no effect. Until time = 400, only allelic values 0 or 1 are allowed at the main locus. This corresponds to Grafen's [45] 'discrete strategy case' for which the predicted equilibrium is the Grafen ESS. After time = 400, new mutations are drawn from a uniform distribution between 0 and 1, with allelic values interpreted as probabilities to cooperate. This resembles Grafen's 'continuous strategy case' (for which the predicted equilibrium is the standard ESS), except that it makes no assumption of small mutational steps or weak selection. The right panel shows the phenotypic correlation between interacting siblings. Dashed line: Grafen ESS. Stippled line: standard ESS.

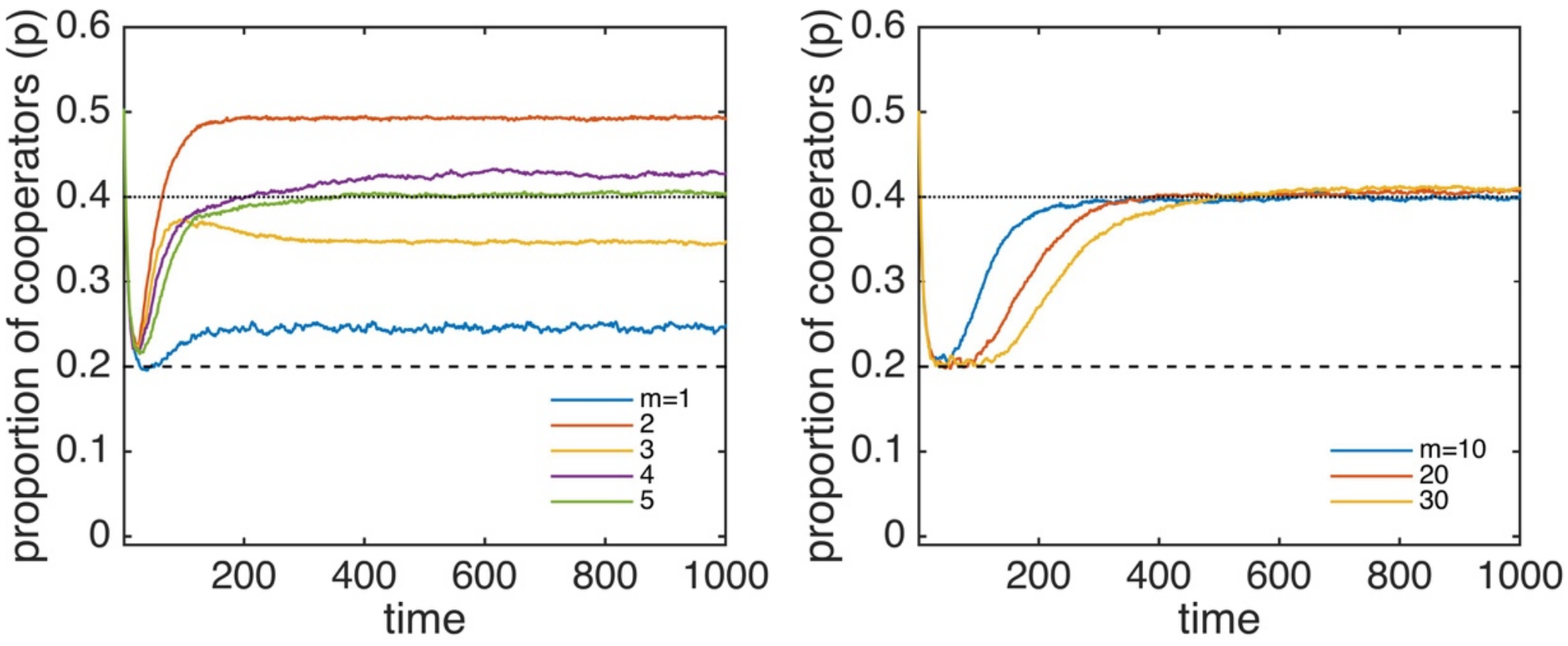

Figure S2 A & B. Main locus with fixed modifiability (i.e., $M = 1$) and $m$ modifier loci per phenotype. The results are split into two graphs for clarity. Dashed line: Grafen ESS. Stippled line: standard ESS.

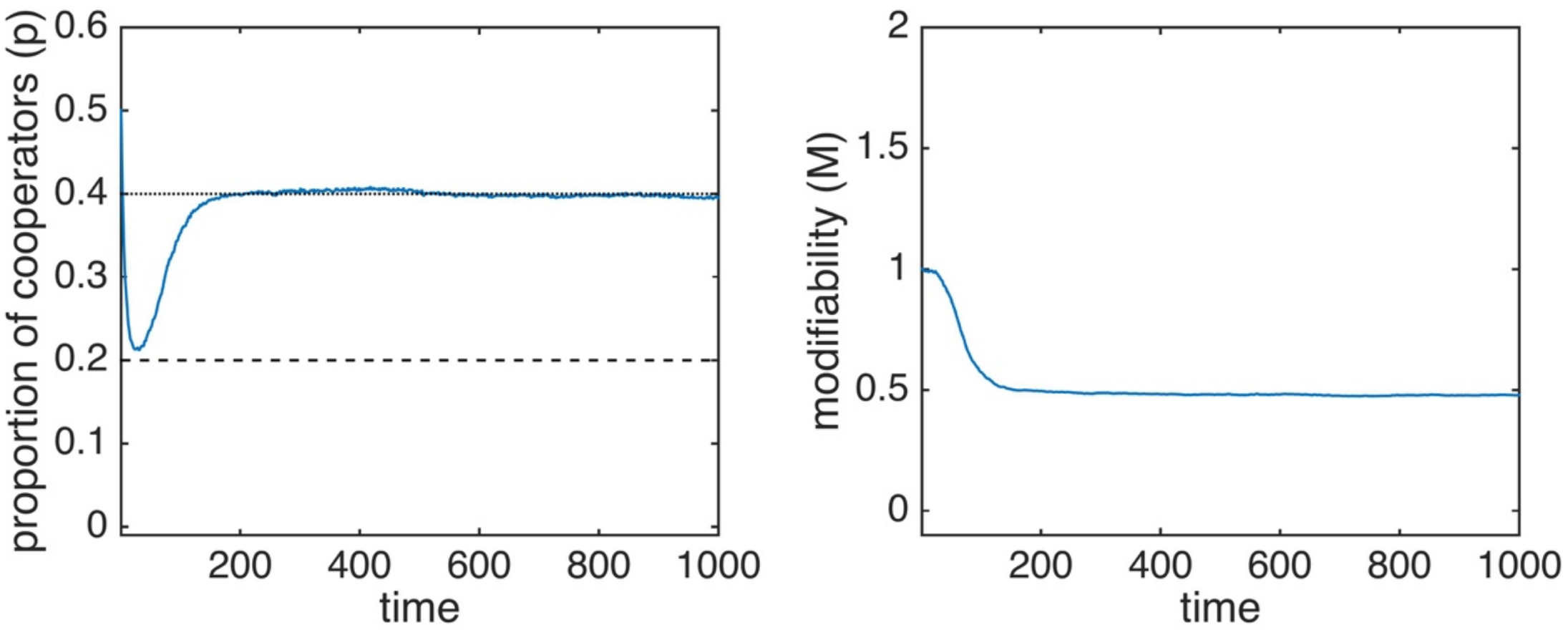

Figure S3. Main locus with evolvable modifiability (i.e. $M$ can vary between 0 and 2) and $m = 1$ modifier locus per phenotype. Dashed line: Grafen ESS. Stippled line: standard ESS.

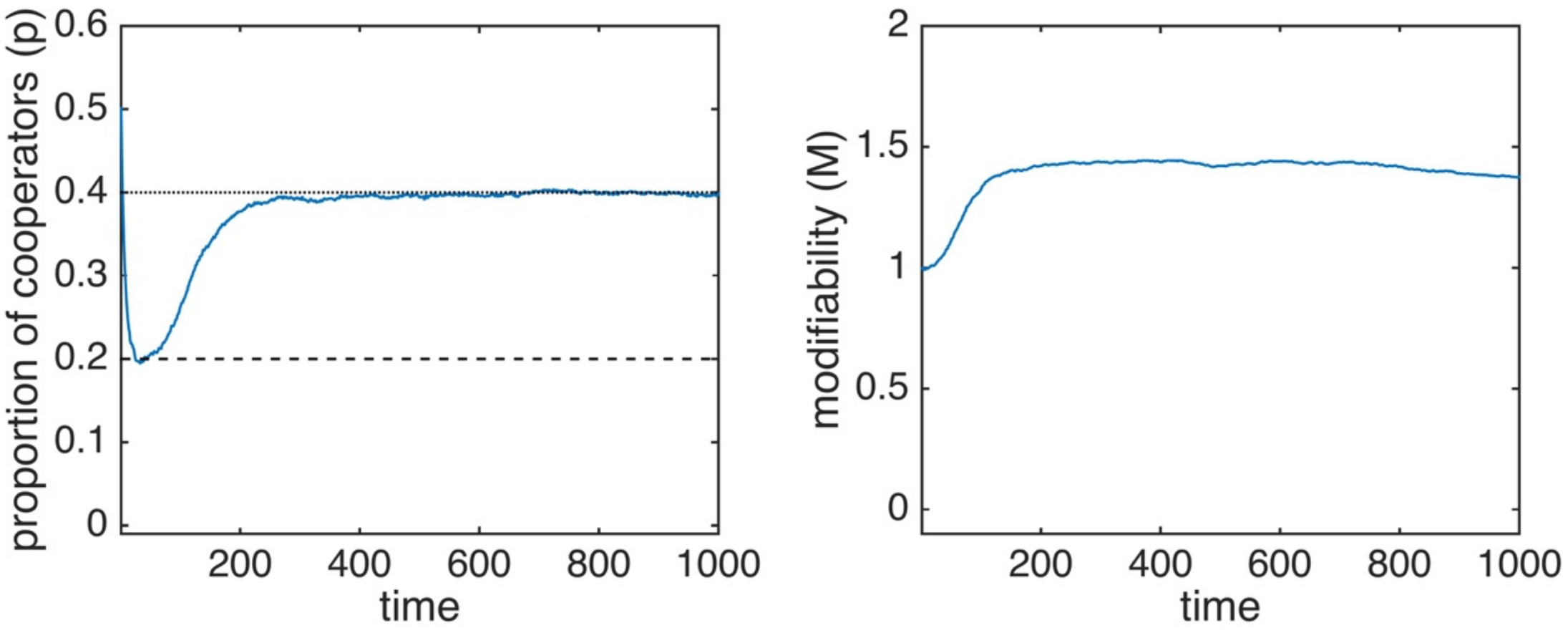

Figure S4. Main locus with evolvable modifiability (i.e. $M$ can vary between 0 and 2) and $m = 10$ modifier loci per phenotype. Dashed line: Grafen ESS. Stippled line: standard ESS.

**Supplementary Material 3: Maximisation of $IF_{\text{folk}}$ vs. $IF_{\text{Hamilton}}$**

Since $IF_{\text{folk}}$ is by definition proportional to the number of reference gene copies produced due to the focal individual's phenotype, it follows that, unless the resident phenotype already maximises $IF_{\text{folk}}$, a rare mutant gene that makes it so (without mirror effect) will spread. This makes maximisation of $IF_{\text{folk}}$ necessary for evolutionary stability. To examine if the same is true for $IF_{\text{Hamilton}}$ (i.e. if, despite their quantitative difference, both are always maximized by the same strategy), we use a simplified version of the model described in Supplementary Material 1. In this simplified version, we consider the special case where the focal individual is certain to face a cooperator – either because cooperation is fixed ($p = 1$), or (in the asymmetric case) because the focal individual adopts role B in a population where the resident strategy is to always cooperate in role A. We then examine whether the behaviour that maximises the focal individual's $IF_{\text{folk}}$ necessarily matches the behaviour that maximises its $IF_{\text{Hamilton}}$. To that end, we consider that the focal individual's behaviour may change independently of the non-focal individual's behaviour – e.g. due to expressing a low-penetrance gene, or due to an experimental intervention. In its given social environment of facing a cooperator, the focal individual obtains

$$IF_{\text{folk}}[\text{cooperate}] = baseline + b + d - c + r(b + d)$$

from cooperating, and

$$IF_{\text{folk}}[\text{defect}] = baseline + b$$

from defecting. Implicit in this formulation is the assumption that the focal individual has no causal effect on whether the non-focal individual incurs the cost of cooperation. (Without this assumption, another indirect component, $-rc$, would need to be included both in $IF_{\text{folk}}[\text{cooperate}]$ and in $IF_{\text{folk}}[\text{defect}]$. This would not affect our conclusion.) Since $IF_{\text{Hamilton}}$ differs from $IF_{\text{folk}}$ in excluding "all components which can be considered as due to the individual's social environment", the components to be excluded here are $(b + d)$ if the focal individual cooperates, or $b$ if the focal individual defects. This yields

$$IF_{\text{Hamilton}}[\text{cooperate}] = baseline - c + r(b + d)$$

from cooperating and

$$IF_{\text{Hamilton}}[\text{defect}] = baseline$$

from defecting. The exclusion of $b$ from both $IF_{\text{Hamilton}}[\text{cooperate}]$ and $IF_{\text{Hamilton}}[\text{defect}]$ is unproblematic in the sense that it does not affect the ranking of the focal individual's options. By contrast, since $d$ occurs in $IF_{\text{folk}}[\text{cooperate}]$ but not in $IF_{\text{folk}}[\text{defect}]$, excluding it to obtain the corresponding $IF_{\text{Hamilton}}$ may affect the ranking of options either in favour of $IF_{\text{Hamilton}}[\text{cooperate}]$ (if $d < 0$) or in favour of $IF_{\text{Hamilton}}[\text{defect}]$ (if $d > 0$). Only when fitness effects are additive ($d = 0$; as originally assumed by Hamilton [6]) can we rely on $IF_{\text{Hamilton}}$ necessarily predicting the same behaviour as $IF_{\text{folk}}$.

However, even when $d \neq 0$, the situation can still be described by a modified version of Hamilton's rule. For example, consider an experimental study in which an individual facing a cooperator is manipulated to cooperate instead of defect. The causal effects of this manipulation are a net loss to the focal individual's reproduction of magnitude $C = -c + d$, and a net gain to the recipient's reproduction of magnitude $B = b + d$. The individual should therefore cooperate if $rB - C > 0$, where $B$ and $C$ are net rather than additive effects. One could call this 'Hamilton's phenotypic rule' as, instead of being concerned with selection for a focal gene, it provides a phenotypic criterion that must be met for a behaviour to increase a focal individual's $IF_{\text{folk}}$.

As another example, consider Creel's paradox as described in Figure 3. Here, the focal individual obtains

$IF_{\text{folk}}[\text{cooperate}] = baseline - c + rb$

from cooperating (i.e., becoming the helper), or

$IF_{\text{folk}}[\text{defect}] = baseline + b$

from defecting (i.e., becoming the breeder), where $baseline = 0$, $c = 0$, $b = 2$ (see Fig 3).

The focal individual should prefer to become the helper if $IF_{\text{folk}}[\text{cooperate}] > IF_{\text{folk}}[\text{defect}]$, yielding

$rb - c > b$,

which (with settings $c = 0$, $b = 2$ as indicated above) predicts correctly that the focal individual should never prefer to be the helper.

Because component $b$ in $IF_{\text{folk}}[\text{defect}]$ reflects help received from the social environment, it should be excluded from $IF_{\text{Hamilton}}$ to obtain

$IF_{\text{Hamilton}}[\text{cooperate}] = baseline - c + rb$

from cooperating and

$IF_{\text{Hamilton}}[\text{defect}] = baseline$

from defecting. Accordingly, $IF_{\text{Hamilton}}[\text{cooperate}] > IF_{\text{Hamilton}}[\text{defect}]$ yields $rb - c > 0$, which (with settings $c = 0$, $b = 2$ as indicated above) predicts wrongly that the focal individual should prefer to be the helper whenever $r > 0$.

As before, the situation can still be described by Hamilton's phenotypic rule (see above). Envisage an experimental study in which a focal breeder is manipulated to become a helper instead (so the former helper becomes the breeder). The causal effects of this manipulation are a net loss to the focal individual's reproduction of magnitude $C = b + c$ (i.e. the $b$ it no longer receives, plus the $c$ it now pays), and a net gain to the non-focal individual's reproduction of magnitude $B = b + c$ (i.e. the $b$ it now receives, plus the $c$ it no longer pays). Accordingly, the focal individual should cooperate (i.e., become the helper) if $rB - C > 0$, where $B$ and $C$ are the net causal effects of its behaviour. This yields $r(b + c) > b + c$, which is not satisfied in the present example - predicting correctly that the focal individual should prefer to remain as the breeder.

*Conclusion*

When fitness effects are non-additive, $IF_{\text{Hamilton}}$ is not in general maximised by the same strategy as $IF_{\text{folk}}$ – implying that $IF_{\text{Hamilton}}$ does not generally work as a phenotypic maximand. However, if organisms maximise $IF_{\text{folk}}$, their behaviour follows a form of Hamilton's rule in which cost and benefit are net effects rather than additive effects. This form of Hamilton's rule corresponds to what Hamilton [1] called the "generalized unrigorous statement of the main principle": "The social behaviour of a species evolves in such a way that in each distinct behaviour-evoking situation the individual will seem to value his neighbours' fitness against his own according to the coefficients of relationship appropriate to that situation."

**Supplementary Material 4: Maximisation of $IF_{\text{folk}}$ vs. neighbour-modulated fitness**
Neighbour-modulated fitness [6] (*NF*) simply counts all of an individual's own offspring, regardless of whether any of them are produced due to the social environment. Because a gene associated with high reproductive success will tend to spread (regardless of the causality of the association), the (average) *NF* of a gene's carriers predicts whether a gene will spread. This is the basis of the so-called 'general form of Hamilton's rule' [32–34], which establishes that *NF* works as an accounting tool for any gene.

Here we ask whether *NF* qualifies as a phenotypic maximand. Consider the model described in Supplementary Material 1, with $d = 0$ for simplicity. In this model, a focal individual that cooperates obtains

$$IF_{\text{folk}} = baseline - c + fb + rb$$

where $f$ is its probability of facing another cooperator (hence receiving a benefit *b*). The corresponding neighbour-modulated fitness is:

$$NF = baseline - c + fb.$$

Note that *NF* excludes the indirect component $rb$ which the focal individual obtains from causing its relative to produce $b$ additional offspring. This leaves the cost of helping, $-c$, as the only causal effect of the focal individual's behaviour that is accounted for in *NF*. Thus, if we envisage an experimental intervention that prevents a costly helping act (with any $b > 0$ and $c > 0$), the focal individual's *NF* would increase as a result. This shows that costly helping is inconsistent with organisms being adapted to maximise their *NF*. In other words, *NF* is not a phenotypic maximand (also see [3]).

Nevertheless, *NF* is highly useful in theoretical models to derive the evolutionarily stable strategy (ESS) of one continuous trait at a time. This is done with the differentiation method of Taylor and Frank [30], which essentially answers the question: "If an organism could choose its genotypic value for a continuous trait *X*, on the assumption that changing its value by 1 unit will be accompanied by a correlated change of *r* units in its relatives, which value should it choose to maximise its neighbour-modulated fitness?" This question is useful to find the ESS because a rare gene affecting *X* will be positively selected if it increases the average *NF* of its carriers. The question does not, however, invoke *NF* as a phenotypic maximand. Whereas a phenotypic maximand should reflect a focal organism's causal effects in its given social environment, the *NF* question invokes a change in the social environment that is merely correlated with, but not caused by, a property of the focal organism.

## Supplementary Material 5: Questions and answers

*Q1 Do you claim that reference genes are a major driving force of phenotypic evolution?*

*Q2 Why should a reference gene be 'rarely or never expressed'?*

*Q3 Does invoking low-penetrance 'reference genes' amount to assuming weak selection, genes of small effect size, and hence (approximately) additive interactions among genes?*

*Q4 By invoking the abstraction and idealization of a reference gene, you reach conclusions about evolutionary dynamics that are merely heuristic. Why not stick with the precision of standard population genetic models?*

*Q5 Why should it matter what outcomes are under an organism's 'control'? In other words, why should we view causality 'through the lens' of a focal organism's effects?*

*Q6 Why should we think of selection in terms of causal effects, rather than genotype-fitness (or trait-fitness) covariance?*

*Q7 Does the mirror effect become unavoidable when a phenotype becomes common, such that organisms increasingly face their own phenotype in their social partners?*

*Q8 Is it plausible that genes without mirror effect exist that affect any given trait?*

*Q9 Does a gene with mirror effect necessarily generate a positive phenotypic correlation between interacting relatives?*

*Q10 Are evolutionary 'end points' always determined by genes without mirror effect, rather than genes with mirror effect?*

*Q11 Can there be an 'evolutionary arms race' between MER genes and genes without mirror effect?*

*Q12 When the "Grafen ESS" is invaded by low-penetrance mutations, could the evolutionary trend towards the standard ESS be halted or reversed if high-penetrance mutations arise at a sufficiently higher rate than low-penetrance mutations?*

*Q13 If the interplay between high- versus low- penetrance genes (unlike between meiotic drive versus Mendelian genes) is really a matter of genetic constraints rather than of conflict, doesn't that make the parliament of genes a misleading metaphor?*

*Q14 Do you have mathematical proof that $IF_{\mathrm{folk}}$ is a phenotypic maximand?*

*Q15 Does calculating a focal gene's causal effect on $IF_{\mathrm{folk}}$ (compared to the counterfactual $\widehat{IF}_{\mathrm{folk}}$) establish the direction of selection on that gene?*

*Q16 Should evolution proceed towards phenotypes with higher $IF_{\mathrm{folk}}$ in a frequency-independent manner?*

*Q17 When counting reference gene copies that come into existence because the focal individual exists, one implicitly invokes the counterfactual possibility that it does not exist. How should this apparently rather far-fetched possibility be interpreted?*

*Q18 Why think of $IF_{\mathrm{folk}}$ as an absolute property of an organism, rather than consider only differences in $IF_{\mathrm{folk}}$ between phenotypes? After all, natural selection works on differences.*

*Q19 Do you predict real organisms to be perfectly adapted to maximise their $IF_{\mathrm{folk}}$?*

*Q20 You define $IF_{\mathrm{folk}}$ so that a reference gene that increases an organism's $IF_{\mathrm{folk}}$ is positively selected by definition. Then you predict that organisms should evolve towards maximising $IF_{\mathrm{folk}}$. Isn't that a circular argument?*

*Q21 According to West & Gardner* [3]*, a phenotypic maximand must be under the organism's 'full control', because organisms can only appear designed to maximise something they can control. Doesn't that exclude* $IF_{\text{folk}}$*, which is in part controlled by others?*

*Q22 Another way to describe the 'double accounting problem' that allegedly plagues* $IF_{\text{folk}}$ *is that "it allows children to be counted many times, as though they had many existences"* [20]*. Is that a valid concern?*

*Q23 With hindsight it seems quite intuitive that organisms should maximise* $IF_{\text{folk}}$*. Has no-one explicitly argued for this before?*

*Q24 If most biologists have intuitively thought about inclusive fitness correctly, does it really matter if we now call it the 'folk definition of inclusive fitness'?*

*Q25 How should empiricists measure* $IF_{\text{folk}}$*?*

*Q26 Is* $IF_{\text{folk}}$ *the unique quantity that qualifies as a phenotypic maximand?*

*Q27 How should we interpret Okasha & Martens'* [25] *result that a quantity they call "Grafen 1979 payoff" qualifies as a phenotypic maximand under broader conditions than* $IF_{\text{SWS}}$*?*

*Q28 Natural selection can be said to maximise a gene's inclusive fitness effect (IFE) in the sense that, at each locus, the allele in a set of possibilities that has the highest IFE should end up being present at equilibrium. Is that the same as* $IF_{\text{folk}}$ *being maximised?*

*Q29 In the formal part of his paper, Hamilton* [6] *defines IF simply as 1 + IFE, where IFE is the focal gene's inclusive fitness effect. This formulation implies that IF is maximised whenever IFE is maximised. Doesn't that contradict the view that there is a deep conceptual distinction between IF (as a property of an organism) and IFE (as a property of a gene or trait)?*

*Q30 Is calculating a focal helping gene's causal effect on* $IF_{\text{folk}}$ *(compared to the counterfactual* $\widehat{IF}_{\text{folk}}[\text{defect}]$*) equivalent to Hamilton's 'stripping procedure'?*

*Q31 The claim that the prescribed 'stripping' of fitness components is problematic for the use of inclusive fitness seems at odds with a lack of theoretical studies that actually involve any 'stripping'. Why?*

*Q32 More often than not, causal effects in biology are probabilistic not deterministic. How do you account for that?*

*Q33 In the statement that "organisms should maximize their expected* $IF_{\text{folk}}$*", over what range of possibilities is the expectation to be taken? Over a given individual's future possibilities, or over all possibilities that a randomly chosen individual at conception may face in its life?*

*Q34 How do you account for so-called 'cancellation effects'* [26,46,47] *due to local competition?*

*Q35 Are you suggesting we should change our methods of modelling social evolution?*

*Q36 What advantage does an 'organismal' view of evolution have over a purely 'gene-centred' view that focusses only on what kinds of genes can be selected for?*

*Q37 Isn't it wrong to assume (as criticised by Dawkins* [20]*) "that an individual organism, as a coherent entity, works on behalf of copies of all the genes inside it"?*

*Q38 Is the individual organism the unique level in the hierarchy of life at which natural selection can be said to optimise performance (i.e., vehicle quality, inclusive fitness)?*

*Q39 Does the kinship theory of genomic imprinting, which posits that evolutionary interests may differ between paternally versus maternally-derived genes, contradict the idea that an individual's genome has a 'majority interest'?*

*Q40 Do you have anything to add for readers who are still sceptical?*

## Reference genes and the parliament of genes

*Q1 Do you claim that reference genes are a major driving force of phenotypic evolution?*
No. We merely claim that phenotypic evolution is largely driven by genes whose phenotypic effects are qualitatively in line with the genome's 'majority interest'. A reference gene is a hypothetical gene whose idealized properties align its evolutionary 'interest' (i.e., the ranking of possible phenotypic options with respect to how well they propagate the gene) with the genome's 'majority interest' as to what phenotype should be expressed. It is a conceptual tool to delineate what phenotypic changes caused by actual genes will increase vehicle quality, implying a high potential to make a lasting contribution to phenotypic design. By contrast, actual genes that match our definition of a reference gene are weak drivers of evolutionary change, being both rare and rarely expressed.

*Q2 Why should a reference gene be 'rarely or never expressed'?*
Penetrance affects the direction of selection because it influences the mirror effect. Genes with or without mirror effect can simultaneously generate selection in opposite phenotypic directions (Figure 2; Supplementary Material 1). Which direction prevails in the long run depends on how genes interact with each other in the cumulative process of multi-locus evolution. Since low penetrance implies weak selection, and genic relatedness matches pedigree relatedness for weakly selected genes over the genome [10], a reference gene 'agrees' with most other genes as to what traits best serve their propagation. A reference gene is therefore representative of the 'evolutionary interest' of the organism. Still, why do we say that a reference gene is 'rarely *or never*' expressed? The reason is that slightly different properties are convenient to use in different contexts: in the context of measuring an organism's vehicle quality, a reference gene is best envisaged as a passive marker to measure the organism's causal effects on gene propagation through mechanisms that apply even to non-expressed genes. (Intuitively, the effect of a trait on the spread of neutral genes is the best measure of the extent to which the trait increases copying of the whole genome.) In the context of evolutionary stability, it is useful to envisage a reference gene as being expressed in a focal organism but not in its social environment, to then ask: what phenotypic changes could the reference gene induce in the organism to propagate more copies of itself (i.e., be positively selected) in the given social environment? Put another way, low penetrance genes are particularly relevant for adaptation because they essentially affect one organism at a time, and hence are in some sense the finest-grained changes that evolution has in its toolbox to optimise organismal design.

*Q3 Does invoking low-penetrance 'reference genes' amount to assuming weak selection, genes of small effect size, and hence (approximately) additive interactions among genes?*
No. As commonly understood, the assumptions of weak selection and small effect size mean that the entire evolutionary process is driven by genes with these properties. By contrast, we do not even assume that low-penetrance genes are common; we merely assume that evolution proceeds *in part* by low-penetrance genes. This makes a crucial difference because, even if evolutionary dynamics are largely driven by high-penetrance genes under strong selection, evolutionary stability needs to be evaluated with regard to mutant genes that can have any degree of penetrance – including low penetrance. From our argument about evolutionary stability (see Q14) this implies that a population cannot be evolutionarily stable unless organisms exhibit phenotypes that maximise $IF_{\text{folk}}$.

*Q4 By invoking the abstraction and idealization of a reference gene, you reach conclusions about evolutionary dynamics that are merely heuristic. Why not stick with the precision of standard population genetic models?*
Our aim is to understand long-term phenotypic evolution. If we hope to understand this multi-locus process by focussing on a single gene (e.g., by adopting the 'gene's eye view' [4]), then that gene's properties should be chosen with that goal in mind - rather than based on what may seem 'typical' for most genes. Faced with a choice between mathematical precision and biological relevance, we think it is better to be approximately right than precisely wrong.

*Q5 Why should it matter what outcomes are under an organism's 'control'? In other words, why should we view causality 'through the lens' of a focal organism's effects?*
The idea of organismal control is intuitively appealing because we all perceive ourselves as coherent agents who pursue certain goals through the changes we cause in the world. Our approach justifies this intuition as follows. Genes are selected to influence each other because their joint effects on the organism they produce mediate their propagation. These co-evolutionary interactions between genes can be metaphorically described as a negotiation process, played out over evolutionary time, about what phenotype should be expressed. Crucially, this process can only shape traits that are at least partly under an organism's control, in that they can be causally influenced by its genes. Such traits include an organism's propensity to help others, but (in the absence of a feasible causal mechanism) not the propensity to receive help, any more than the propensity to make the sun shine. This cumulative co-evolutionary process makes organismal design goal-directed in that an organism's causal effects come to reflect its 'evolutionary interest'.

*Q6 Why should we think of selection in terms of causal effects, rather than genotype-fitness (or trait-fitness) covariance?*
We are motivated by an interest in long-term outcomes – which requires complementing the familiar guiding question: "*what kind of gene will be positively selected?"* by adding *"..., such that its phenotypic effect is not eliminated in the long run*". Genotype-fitness covariance is crucial to predict short-term change, but it draws no distinction between selection for rogue and adaptive genes. Similarly, where trait-fitness covariance does not reflect a causal relationship, it can be misleading because it predicts the short-term spreading of maladaptive traits (e.g. as a pleiotropic effect of an otherwise useful gene) that are not maintained in the long run because negative pleiotropic effect can often be modified by evolution at other loci.

**The mirror effect**

*Q7 Does the mirror effect become unavoidable when a phenotype becomes common, such that organisms increasingly face their own phenotype in their social partners?*
No. As defined here, the mirror effect is a property of a gene, not of a phenotype or organism. Even if phenotypes are almost completely uniform, a gene making a small difference to the phenotype may or may not be simultaneously expressed in interacting relatives.

*Q8 Is it plausible that genes without mirror effect exist that affect any given trait?*
For our argument to hold, it only needs to be the case that they can (and eventually will) arise in the long run, even in systems where they don't presently exist. We believe that this is a plausible assumption. In general, genes can have any level of penetrance, from 100% to 0%. As we go from high to low penetrance, the mirror effect weakens and eventually becomes negligible. For example, if a gene is expressed in only 1% of its carriers, an individual expressing it will almost exclusively interact with social partners who do not express it. Alternatively, the mirror effect can be avoided by genes being expressed conditional on some (perhaps arbitrary) asymmetry between individuals.

While theoretical models often exclude conditionality *a priori*, it is worth noting that conditionality in nature should tend to evolve precisely in those circumstances in which the mirror effect determines whether a trait is selected for or against; i.e. when individuals can gain from unilaterally changing their strategy. An impressive example of natural selection's power to overcome genetic constraints due to the mirror effect is a multicellular body, in which genetically identical cells do very different things.

*Q9 Does a gene with mirror effect necessarily generate a positive phenotypic correlation between interacting relatives?*
No. For example, consider a gene with penetrance $P = 0.5$ that induces helping in symmetric pairwise interactions between relatives. This gene has a moderately strong mirror effect: if both individuals have it and the focal individual expresses it, then the non-focal individual also expresses it with 50% probability. Nevertheless, if the gene is fixed in the population, observing a focal individual's behaviour reveals no information (beyond the base rate of helping) about the non-focal individual's likely behaviour. Despite behavioural tendencies being perfectly matched, no positive correlation arises at the level of actual behaviour. This example illustrates that genes with very low penetrance ('without mirror effect') are not always needed to overcome a disadvantageous tendency of organisms to disproportionally face their own phenotype.

*Q10 Are evolutionary 'end points' always determined by genes without mirror effect, rather than genes with mirror effect?*
No. Sometimes genes with mirror effect invade more easily than those without; sometimes the reverse is true (Supplementary Material 1). In both cases, the possible invasion is relevant for what population state qualifies as an equilibrium. In that sense, both kinds of genes influence the 'end point'. $IF_{\text{folk}}$, however, is maximised in either case. Why? Because otherwise it would not be an 'end point', as a reference gene that increases $IF_{\text{folk}}$ could still invade. Moreover, the view that genes without mirror effect are particularly important may be justified by the finding that, under some conditions (namely, synergy and additivity), genes with or without mirror effect never 'disagree' about what traits are selected for and there are no MER genes (Supplementary Material 1). Hence $IF_{\text{folk}}$ will end up being maximized even when evolution is entirely driven by genes with mirror effect. And under the remaining conditions (interference), genes without mirror effect shape the equilibrium because they can invade more easily.

**Mirror effect rogue genes**

*Q11 Can there be an 'evolutionary arms race' between MER genes and genes without mirror effect?*
No. MER genes are under no selection to resist having their penetrance modified to reduce the mirror effect. For example, assume that the full-penetrance defector genes in Fig. 2 come in two variants: one that is prone to have its penetrance slightly reduced by a modifier gene, and another that resists such modification. Then individuals simultaneously possessing both the modifier gene and the modification-prone gene behave, in effect, as if possessing a low-penetrance cooperator gene that (sometimes) allows them to reap the benefits of unilateral cooperation. This generates selection for proneness to modification, not against it.

*Q12 When the "Grafen ESS" is invaded by low-penetrance mutations, could the evolutionary trend towards the standard ESS be halted or reversed if high-penetrance mutations arise at a sufficiently higher rate than low-penetrance mutations?*
No. The (relative) frequency of low-penetrance mutations should affect the speed, but not the general direction of evolution towards the standard ESS. To see why, consider an initially uncooperative population that is invaded by a full-penetrance cooperator gene. As the cooperator gene spreads, cooperators increasingly face other cooperators, and become more prone to experiencing interference between matching phenotypes (recall that $d < 0$ is a requirement for the Grafen ESS to exist). This tendency to suffer from interference is exacerbated by the mirror effect, which eventually stops the spread of cooperation at the Grafen ESS. Genes without mirror effect invade the Grafen ESS precisely because they enable their carriers to avoid (and, indeed, reverse) the disadvantageous tendency to disproportionally face their own type. When genes without mirror effect invade, they therefore weaken the phenotypic correlation in the population. As long as the correlation remains positive, individuals that can 'escape' the correlation (i.e., that are freed from the disadvantageous tendency of disproportionally facing their own phenotype) are better off. Hence low-penetrance genes that induce a switch in their carriers' phenotype continue to enjoy a selective advantage. An end point is only reached once the phenotypic correlation in the population is zero, which occurs at the standard ESS. Can a high rate of high-penetrance mutations undermine this process? No, for the following reason: once the phenotypic correlation $R$ has weakened (i.e., $R < r$), a newly mutated full-penetrance gene (whose phenotypic effect must override all other genes to achieve full penetrance) will disadvantage its carriers compared to 'resident' individuals of the same phenotype. This is so because carriers of a mutant full-penetrance gene suffer more than others from the disadvantageous tendency to face their own phenotype. This prevents (re-) invasion of full-penetrance genes. Our simulation results illustrate this principle (Supplementary Material 2).

*Q13 If the interplay between high- versus low- penetrance genes (unlike between meiotic drive versus Mendelian genes) is really a matter of genetic constraints rather than of conflict, doesn't that make the parliament of genes a misleading metaphor?*
In general, we find the metaphor apt for multi-locus evolution because it captures the idea that, although various genes may (for whatever reason) pull phenotypic evolution in opposing directions in the short term, we can still predict the likely long-term outcomes of adaptation when we consider the combined phenotypic effects of many genes. Nevertheless, we concede that the numerical imbalance of genes implied by the metaphor is not equally crucial for countering the effects of all rogue genes. On the one hand, since meiotic drive genes can always invade anew given appropriate mutations, they can fuel endless cycles of invasion and counter-selection in which the majority of the genome's numerical preponderance should play a crucial role. By contrast, since MER genes can no longer invade once a population has reached a phenotypic equilibrium (Q12), they do not need to be continuously kept in check in that way.

**The folk definition of inclusive fitness**

*Q14 Do you have mathematical proof that $IF_{\text{folk}}$ is a phenotypic maximand?*
No, but we have a logical proof that requires no formal mathematics. It is summarised by the "argument about evolutionary stability" of section 6. It is a proof by contradiction that rests on the incompatibility of three premises:
(1) The population is phenotypically stable, such that no rare mutant gene can be positively selected that encodes a strategy other than the 'resident' strategy currently adopted by the majority of organisms.
(2) Each strategy in the strategy set can be encoded by genes with any degree of penetrance (including low penetrance); and all feasible mutations arise in the long run.
(3) Organisms adopting the resident strategy do *not* behave as if to maximise their $IF_{\text{folk}}$.

We begin by noting that the statement "the organism behaves as if to maximise its $IF_{\text{folk}}$" is equivalent to the statement "the organism behaves as if to maximise the propagation of a rare, low-penetrance gene". This equivalence stems from the definition of *vehicle quality* as the sum of an organism's causal effects on the propagation of a low-penetrance gene, and from vehicle quality being proportional to $IF_{\text{folk}}$. Next, envisage a mutation of a previously neutral, low-penetrance gene, which, when expressed, changes the focal organism's strategy so that it behaves as if to maximise the propagation of a low-penetrance gene. Premises (2) and (3) ensure that such a mutation will eventually arise. Since it is true by assumption that the mutant gene did something to improve its propagation success, it must be positively selected (however weakly). And since the mutated gene achieved this by inducing a phenotypic change, premise 1 is violated. Hence premises (1) - (3) cannot be met simultaneously.

*Q15 Does calculating a focal gene's causal effect on* $\text{IF}_{\text{folk}}$ *(compared to the counterfactual* $\widehat{\text{IF}}_{\text{folk}}$*) establish the direction of selection on that gene?*

No. Positive selection for a (weakly selected, Mendelian) gene can be inferred if expressing it increases the focal organism's $IF_{\text{folk}}$, but rogue genes that reduce $IF_{\text{folk}}$ can also be selected for. We illustrate this with an example based on the model of Supplementary Material 1. A rare full-penetrance cooperator gene is positively selected if

$r(b + d) > c$ (7).

On the other hand, $IF_{\text{folk}}[\text{cooperate}] - \widehat{IF}_{\text{folk}}[\text{defect}] > 0$ yields

$r(b + d + rd) > c$ (8)

as the condition where expressing the focal gene causally increases $IF_{\text{folk}}$. Here, $IF_{\text{folk}}[\text{cooperate}] = \, baseline - c + f(b + d) + r(b + fd)$; $\widehat{IF}_{\text{folk}}[\text{defect}] = baseline + fb$; and *f* is replaced by *r* (because *f* matches *r* for a rare full-penetrance gene). Given synergy or additivity ($d \geq 0$), condition (8) is always met when condition (7) is met, meaning that expressing a positively selected gene increases $IF_{\text{folk}}$. Given interference ($d < 0$), however, condition (7) can be met even while $IF_{\text{folk}}[\text{cooperate}] - \widehat{IF}_{\text{folk}}[\text{defect}] < 0$, i.e. while

$r(b + d + rd) < c$ (9).

When conditions (7) and (9) hold simultaneously, the focal gene is selected for despite its expression decreasing $IF_{\text{folk}}$. In short, the gene is a *mirror effect rogue gene* that opposes the evolutionary trend towards increased $IF_{\text{folk}}$. This situation, however, generates selection for any low-penetrance modifier gene that would prevent the focal cooperator gene from being expressed. Such a modifier, when expressed, will cause the focal organism to have $\widehat{IF}_{\text{folk}}[\text{defect}]$ instead of $IF_{\text{folk}}[\text{cooperate}]$. The resultant change is positive (i.e. $\widehat{IF}_{\text{folk}}[\text{defect}] - IF_{\text{folk}}[\text{cooperate}] > 0$, implying selection for the modifier) whenever condition (9) holds; i.e., whenever expressing the cooperator gene reduces $IF_{\text{folk}}$ in the first place. This illustrates the principle that genes which reduce $IF_{\text{folk}}$ face counter-selection in the long run.

*Q16 Should evolution proceed towards phenotypes with higher* $\text{IF}_{\text{folk}}$ *in a frequency-independent manner?*

No. $IF_{\text{folk}}$ is evaluated *in a given (social) environment*, which changes as phenotype frequencies change. So, a general trend towards phenotypes with higher $IF_{\text{folk}}$ is fully compatible with frequency-dependent selection (see Supplementary Material 1 for examples). Although a frequency-dependent trait might either increase or decrease $IF_{\text{folk}}$ in different circumstances, this does not preclude (barring rogue genes) a selective trend at each point in time towards phenotypes yielding higher $IF_{\text{folk}}$.

*Q17 When counting reference gene copies that come into existence because the focal individual exists, one implicitly invokes the counterfactual possibility that it does not exist. How should this apparently rather far-fetched possibility be interpreted?*

To predict which phenotype should evolve among a set of alternatives, it is sufficient to consider the differences in $IF_{\text{folk}}$ among the immediate alternatives. For example, to check if an action performed at time *t* increases an organism's $IF_{\text{folk}}$, all causal effects of the organism's existence up to time *t* can be taken as given. This does not require measuring $IF_{\text{folk}}$ in absolute terms, and so does not invoke the idea of a focal organism's non-existence. Nevertheless, to preserve the full power of Hamilton's idea that organisms are adapted to maximise their *IF*, it seems desirable that *IF* also be measurable in absolute terms (at least in principle; but see Q25), which is why we invoke 'non-existence' as a reference point. The biological justification is as follows: for a freshly conceived embryo, it should be mechanistically possible to self-abort instead of develop. And if the chances of success of continued development are slim, this could be an adaptive strategy. For example, if an embryo detects internal physiological clues that indicate major developmental problems, self-abortion could allow its mother to have another offspring sooner. The inclusive fitness outcome of immediate self-abortion thus sets a baseline (i.e., $IF_{\text{folk}} = 0$), which an organism must improve upon to make its continued existence adaptive. This is a variation on Dawkins' [20] view: "We could compare the effects of his choosing to perform act X rather than act Y. Or we could take the effects of his lifetime's set of deeds and compare them with a hypothetical lifetime of total inaction – as though he had never been conceived. It is this latter usage that is normally meant by the inclusive fitness of an individual organism."

*Q18 Why think of $IF_{\text{folk}}$ as an absolute property of an organism, rather than consider only differences in $IF_{\text{folk}}$ between phenotypes? After all, natural selection works on differences.*
This is partly a matter of taste. We have heard this objection from theoreticians, whereas empirically minded people readily embrace the 'absolute property' viewpoint. To see why, consider a hypothetical non-social species for which $IF_{\text{folk}}$ is simply a count of the number of offspring an individual produces. Should we tell biologists studying this species that counting offspring is meaningless, because all that matters are differences? That could be seen as confusing because absolute quantities must exist before differences can be calculated. If we wish to understand this hypothetical species' adaptations, our working hypothesis should be that any putatively adaptive trait increases its bearer's number of offspring. This makes absolute offspring number a meaningful quantity. If we wish to invoke organismal adaptation as an optimising force, we need to articulate what is being optimised.

*Q19 Do you predict real organisms to be perfectly adapted to maximise their $IF_{\text{folk}}$?*
No. We don't expect perfection in nature. But the notion of an 'optimal' phenotype is useful to generate testable predictions, and as a reference to distinguish adaptive from non-adaptive traits. Nor do we claim that an optimality approach captures all interesting biological phenomena. But whenever we wish to use optimality, which is commonplace in evolutionary ecology, we must be prepared to answer the question: "optimal for what"? Our proposed answer is: "optimal for maximising the organism's $IF_{\text{folk}}$". And we are unaware of a similarly general alternative answer.

*Q20 You define $IF_{\text{folk}}$ so that a reference gene that increases an organism's $IF_{\text{folk}}$ is positively selected by definition. Then you predict that organisms should evolve towards maximising $IF_{\text{folk}}$. Isn't that a circular argument?*
No. If the argument were circular, the prediction would necessarily, albeit trivially, be true. This is not the case: at least in principle, our prediction might not fit patterns in the real world, depending on, among other things, mechanistic constraints limiting adaptation; environmental changes limiting the fit between organisms and their environment; and, most importantly, the (approximate) truth of our postulate that reference genes are useful for predicting phenotypic evolution.

*Q21 According to West & Gardner* [3]*, a phenotypic maximand must be under the organism's 'full control', because organisms can only appear designed to maximise something they can control. Doesn't that exclude $IF_{\text{folk}}$, which is in part controlled by others?*
The requirement of 'full', if taken to mean exclusive, control is unnecessary because a consequence can have several causes. For example, if A and B must both occur to bring about C, then an organism that controls A can have a causal effect on C whenever B occurs - regardless of its control of B. Nevertheless, we agree with West and Gardner's main conclusion about 'control': namely, that neighbour-modulated fitness does not qualify as a phenotypic maximand (see Supplementary Material 4).

*Q22 Another way to describe the 'double accounting problem' that allegedly plagues $IF_{\text{folk}}$ is that "it allows children to be counted many times, as though they had many existences"* [20]*. Is that a valid concern?*
No. Since a given causal effect may have several causes (see Q21), summing each individual organism's causal effects on reproduction need not equal the population's total reproduction. For example, if organisms A and B must cooperate to jointly produce *n* offspring, then A's decision to cooperate causes *n* (instead of 0) offspring to exist. And so too does B's decision to cooperate. Yet these statements do not conflict with the premise that the total number of offspring is *n*, not 2*n.* The key point is that considering one cause at a time does not entail a commitment to adding up the consequences of different causes. While it is meaningful to evaluate a focal organism's $IF_{\text{folk}}$ in a given environment (i.e., keeping all but the focal organisms' phenotype constant), it is not meaningful to add up the inclusive fitnesses of all population members. To be sure, the above argument does not directly speak to the validity of our claim that $IF_{\text{folk}}$ is a phenotypic maximand. Instead, it merely serves to show that $IF_{\text{folk}}$ is defined free of this alleged logical contradiction.

*Q23 With hindsight it seems quite intuitive that organisms should maximise $IF_{\text{folk}}$. Has no-one explicitly argued for this before?*
Not that we are aware of. Queller [13] came close to our conclusion in his discussion of Creel's paradox. But he stopped short because he treated inclusive fitness as an accounting tool for a focal gene, not as a phenotypic maximand of an organism. In particularly, he recognised that a gene 'for' becoming the breeder is favoured because, when considering selection at a focal locus, there is no reason to 'strip' the effects of genes at other loci. He also noted that selection for preferring to become the breeder is inconsistent with maximisation of $IF_{\text{Hamilton}}$. But he did not say what is being maximised instead.

*Q24 If most biologists have intuitively thought about inclusive fitness correctly, does it really matter if we now call it the 'folk definition of inclusive fitness'?*
No. We hope that $IF_{\text{folk}}$ will become known as 'inclusive fitness', whereas $IF_{\text{Hamilton}}$ will become a historical footnote. We believe this would be consistent with Hamilton's original motivation for coining the term: to generalise the classic idea of 'fitness' as the property of an organism which natural selection tends to maximise. This idea has been extremely powerful and continues to play a central role in evolutionary explanations that make sense to broad audiences, rather than only to mathematically inclined specialists. It has motivated countless empirical studies that attempt to quantify selection on social behaviours. In part, our motivation for revisiting the definition of inclusive fitness has been to ensure agreement between the ways that empiricists and theoreticians explain why evolution had led to certain types of traits predominating in nature.

*Q25 How should empiricists measure $IF_{\text{folk}}$?*
We see little reason to attempt to measure $IF_{\text{folk}}$ in absolute terms. To test whether a social trait increases an organism's $IF_{\text{folk}}$, one should test whether the trait's causal effects meet Hamilton's phenotypic rule (supplementary material 3). Ideally, one should measure these effects experimentally, by comparing manipulated individuals with control individuals inhabiting closely matched social environments. Then, all systematic differences between treatments with respect to the reproductive success of the focal organism and its relatives can be causally attributed to the focal trait. Crucially, in contrast to tests inspired by $IF_{\text{Hamilton}}$ (review: [48]), there is no need to assume that traits have additive effects. We emphasise that, since $IF_{\text{folk}}$ should be measured in a *given* environment, one should not manipulate multiple individuals that directly interact with each other. For example, by forcing two interacting individuals to cooperate in a prisoner's dilemma, one would increase the $IF_{\text{folk}}$ of both. But this manipulation would be uninformative as to whether cooperating (instead of defecting) increases each individual organism's $IF_{\text{folk}}$.

*Q26 Is $IF_{\text{folk}}$ the unique quantity that qualifies as a phenotypic maximand?*
No. As noted by Okasha & Martens [25], for any given function that qualifies as a phenotypic maximand there exist infinitely many transformations that also qualify. So if $IF_{\text{folk}}$ is a maximand, then so is any function $F = IF_{\text{folk}} + x$, where $x$ is a constant on which the focal organism has no causal effect. For example, if all "effects due to the social environment" can be deemed beyond a focal organism's control, then denoting these as $-x$ recovers $IF_{\text{Hamilton}}$ from $IF_{\text{folk}}$. Similarly, letting $x$ be the reproduction of relatives that is unaffected by the focal organism recovers the 'simple weighted sum' definition of inclusive fitness ($IF_{\text{SWS}}$), which adds up the reproduction of a focal organism and all its relatives, weighted by relatedness. Although Grafen [16] rejected $IF_{\text{SWS}}$, Okasha & Martens [25] found that $IF_{\text{SWS}}$ in fact qualifies as a phenotypic maximand, at least under the same conditions as $IF_{\text{Hamilton}}$. Moreover, the above argument implies that $IF_{\text{SWS}}$ qualifies under even broader conditions, namely under the same conditions as $IF_{\text{folk}}$. Nevertheless, we advise against using $IF_{\text{SWS}}$ because it diverts attention from the causal processes that matter. For example, any comparison of $IF_{\text{SWS}}$ between individuals differing in their number of relatives is confounded by the latter.

*Q27 How should we interpret Okasha & Martens'* [25] *result that a quantity they call "Grafen 1979 payoff" qualifies as a phenotypic maximand under broader conditions than $IF_{\text{SWS}}$?*
This result reflects the restrictive assumption that genes with incomplete penetrance do not exist. The "Grafen 1979 payoff" function is given by $U(i,j) = rV(i,i) + (1-r)V(i,j)$, where $V(i,i)$ is a focal individual's (hypothetical) payoff if it were to play against its own strategy; $V(i,j)$ is its actual payoff playing against a non-focal individual; and $r$ is relatedness. Okasha & Martens' result states that, at evolutionary equilibrium, each individual should behave as if to maximise $U$. The rationale is as follows. Consider a rare mutant gene that is always expressed and that fully specifies an individual's strategy. Carriers of this mutant gene experience payoff $V(i,i)$ with probability $r$, and $V(i,j)$ with probability (1 - $r$). Thus, since the "Grafen 1979 payoff" equals the expected payoff (neighbour-modulated fitness) of a mutant gene's carriers, there is scope for a mutant gene to invade (i.e., to increase the reproductive success of its carriers) whenever the population is not in a state where its members maximise $U$. But this rationale hinges on the assumption that strategies are fully specified by a single gene that is always expressed. Without that assumption, it is not the case that carriers of a rare mutant gene face their own strategy with probability $r$. Indeed, we show in Supplementary Materials 1& 2 that the strategy (i.e., the 'Grafen ESS') which maximises $U$ is unstable when genes with incomplete penetrance exist.

*Q28 Natural selection can be said to maximise a gene's inclusive fitness effect (IFE) in the sense that, at each locus, the allele in a set of possible alleles that has the highest IFE should end up being present at equilibrium. Is that the same as $IF_{\text{folk}}$ being maximised?*
No. Selection at a given locus, with a given set of alleles, does not imply maximising-behaviour of individuals (see Fig. 2). Hence the meaning of the phrase "inclusive fitness is maximised" is obscured whenever authors fail to distinguish between a gene's *IFE* and an individual's *IF* .

**Hamilton's inclusive fitness**

*Q29 In the formal part of his paper, Hamilton* [6] *defines IF simply as 1 + IFE, where IFE is the focal gene's inclusive fitness effect. This formulation implies that IF is maximised whenever IFE is maximised. Doesn't that contradict the view that there is a deep conceptual distinction between IF (as a property of an organism) and IFE (as a property of a gene or trait)?*
No. Hamilton used this definition in a specific model in which baseline fitness was 1 and all social effects were attributable to a single gene. Hence, in this special case, it makes no difference whether *IF* is thought of as capturing the causal effects of an entire organism or a single gene. In general, however, Hamilton stated repeatedly (e.g. see the quote in our introduction) that he intended *IF* to capture the causal effects of an organism. It is worth adding that, regardless of Hamilton's original intentions, we intend $IF_{\text{folk}}$ to capture the causal effects of an organism as a whole.

*Q30 Is calculating a focal helping gene's causal effect on* $IF_{\text{folk}}$ *(compared to the counterfactual* $\widehat{IF}_{\text{folk}}[\text{defect}]$*) equivalent to Hamilton's 'stripping procedure'?*
No. If a focal cooperator receives benefit *b* + *d* from another cooperator (see Supplementary Material 3), then subtracting $\widehat{IF}_{\text{folk}}[\text{defect}]$ excludes only the *b* but not the *d* from the resultant causal effect. This is because only the *b* is received irrespective of the focal individual's phenotype. By contrast, according to Hamilton, "*all* components [i.e., both *b* and *d*] which can be considered as due to the individual's social environment" should be excluded. We note that $IF_{\text{Hamilton}}$ has the unusual property of being mathematically designed to exclude components that are attributable to non-focal causes. Since $IF_{\text{folk}}$ does not share this property, it behaves more like variables with which empirical biologists are familiar. For example, when we state that a fertilizer has effect *x* on plant height, we are not tempted to re-define plant height to exclude components caused by other factors like the rainfall. Instead, we avoid confounding factors by performing controlled experiments to isolate the effect of the fertilizer: i.e., we measure the fertiliser's effect in a given environment. Likewise, causal effects on $IF_{\text{folk}}$ should be measured in a given environment.

*Q31 The claim that the prescribed 'stripping' of fitness components is problematic for the use of inclusive fitness seems at odds with a lack of theoretical studies that involve 'stripping'. Why?*
Because the 'stripping' is not done explicitly – not even in Hamilton's original study [6]. Instead, in practice, double accounting is avoided by using neighbour-modulated fitness. The 'stripping' then happens implicitly, when an expression for neighbour-modulated fitness is re-interpreted as inclusive fitness. Specifically, re-interpreting *rb* as a benefit provided (rather than received) results in an expression that contains no received benefits, so that it appears to have been stripped of them. This 'implicit stripping' is unproblematic in that an inclusive fitness effect, once correctly calculated from neighbour-modulated fitness, correctly predicts selection regardless of its verbal interpretation. By contrast, the explicit 'stripping' prescribed in Hamilton's definition of $IF_{\text{Hamilton}}$ reflects the (unjustified) idea that, to be consistent, the same principle (i.e., that effects due to the social environment be 'stripped') must apply equally to effects of whole organisms and of individual genes.

**Scope of the theory**

*Q32 More often than not, causal effects in biology are probabilistic not deterministic. How do you account for that?*

By causal effects of an organism we mean expected (arithmetic mean) effects. Thus, when we say that organisms are selected to maximize $IF_{\text{folk}}$, we really mean their *expected* $IF_{\text{folk}}$. Perhaps surprisingly, this is appropriate even in fluctuating environments, which are often said to select for genotypes with high *geometric* mean success ('bet-hedging') [49,50]. The apparent contradiction between arithmetic and geometric mean success disappears once one realizes that an evolutionarily relevant measure of reproductive success must account for offspring reproductive value: what matters is not just the number of offspring, but the sum of their reproductive value [8]. In a fluctuating environment, each offspring's reproductive value depends on the context: in a bad year, where population size is small, each offspring is more valuable (i.e., it is a bigger share of the population) than in a good year, where population size is large. Once this is accounted for, organisms are selected to maximize *arithmetic* mean success even in fluctuating environments [8]. In our argument, the possibility of fluctuating environments (and of sustained positive or negative population growth) is implicitly accounted for by weighting offspring by reproductive value.

*Q33 In the statement that "organisms should maximize their expected* $IF_{\text{folk}}$*", over what range of possibilities is the expectation to be taken? Over a given individual's future possibilities, or over all possibilities that a randomly chosen individual at conception may face in its life?*
Both types of expectation should be maximized – at least insofar as traits are concerned that can be modified throughout life at low cost (e.g. behaviour). On the one hand, natural selection shapes phenotypic strategies that specify what phenotype to exhibit in given circumstances. Each organism is endowed with such a (genetically specified) strategy at conception. If one strategy consistently outperforms its alternatives (i.e., yields higher expected $IF_{\text{folk}}$ at conception), it will spread because genes contributing to it enjoy higher propagation success. Ultimately, this makes expected $IF_{\text{folk}}$ at conception the relevant criterion for selection. On the other hand, a strategy's expected $IF_{\text{folk}}$ at conception is maximised if each organism adopting it behaves optimally in its local circumstances (i.e., if, within its limits in perception and flexibility, it maximizes its own expected $IF_{\text{folk}}$). Analogous to standard (non-social) theory of dynamic optimisation [51], organisms should therefore behave at all times as if to maximise the portion of their expected $IF_{\text{folk}}$ that is still in the future (i.e., their *inclusive reproductive value*).

*Q34 How do you account for so-called 'cancellation effects'* [26,46,47] *due to local competition?*
Cancellation effects arise when relatives compete for limited reproductive opportunities. For example, helping your sister to produce a niece is not a good way to propagate your genes if the niece subsequently competes with your daughter for a single breeding opportunity. In our definition of vehicle quality, this is implicitly accounted for when offspring are weighted by their reproductive value: if the niece and daughter each have a 50% chance of winning the single breeding opportunity, then the niece's existence reduces the daughter's reproductive value by half.

*Q35 Are you suggesting we should change our methods of modelling social evolution?*
Not necessarily. In particular, our theory is consistent with the Taylor-Frank method and the corresponding 'inclusive fitness method' of Taylor et al. [31]. We note, however, that these methods are conceptually based on the idea of an accounting tool for genes of small effect size only, whereas one can directly invoke $IF_{\text{folk}}$ as a phenotypic maximand. Indeed, this latter option was suggested long ago by Maynard Smith [52] for game-theory and optimisation models. Concerns about this method by Grafen [24] led to its demise, but they are based on misleading results caused by *mirror effect rogue genes*. If we are only interested in phenotypic outcomes with long-term stability [28], these concerns disappear (Fig. 2, Supplementary Material 1-2). Invoking $IF_{\text{folk}}$ as a maximand has the practical advantage of allowing a broader range of optimisation techniques to be used (e.g., dynamic programming [53]), which – unlike the Taylor-Frank method – are not limited to (locally) optimising one continuous trait at a time.

**Levels of selection**

*Q36 What advantage does an 'organismal' view of evolution have over a purely 'gene-centred' view that focusses only on what kinds of genes can be selected for?*
We think a good part of Darwin's key insight about the design-like appearance of organisms remains unaccounted for if we consider only one gene at a time. Gene-level theories do not, by themselves, explain complex organismal design. They need to be complemented with a higher-level principle that tends to lead phenotypic contributions of individual genes in a coherent direction: namely, towards better-adapted organisms. Without such a principle, complex organismal design has to be regarded as a purely incidental – and hence ultimately unexplained – by-product of gene-level selection. Dawkins [20] came close to acknowledging this when he wrote: "Fundamentally, what is going on is that replicating molecules ensure their survival by means of phenotypic effects on the world. It is only incidentally true that those phenotypic effects happen to be packaged up into units called individual organisms. We do not at present appreciate the organism for the remarkable phenomenon it is. We are accustomed to asking, of any widespread biological phenomenon, 'What is its survival value?' But we do not say, 'What is the survival value of packaging life up into discrete units called organisms?' We accept it as a given feature of the way life is. […] I am not necessarily objecting to this focus of attention on individual organisms, merely calling attention to it as something that we take for granted. Perhaps we should stop taking it for granted and start wondering about the individual organism, as something that needs explaining in its own right, just as we found sexual reproduction to be something that needs explaining in its own right."

*Q37 Isn't it wrong to assume (as criticised by Dawkins* [20]*) "that an individual organism, as a coherent entity, works on behalf of copies of all the genes inside it"?*
Not necessarily. It will still be the case that organisms usually work as coherent entities, if long-term evolution tends to shape organisms that happen to act in such a way. In the words of Mayr [54], "When entities are combined at a higher level of integration, not all the properties of the new entity are necessarily a logical or predictable consequence of the properties of the components." Invoking coherent evolutionary interests of individuals is a useful heuristic to the extent that neglected details (e.g., the occurrence of rogue genes) tend not to have lasting effects on organismal design.

*Q38 Is the individual organism the unique level in the hierarchy of life at which natural selection can be said to optimise performance (i.e., vehicle quality, inclusive fitness)?*
No. In principle, the concept of vehicle quality can be applied at any level of biological organization. One can always ask: what characteristics make entity X a good vehicle for gene propagation? For example, in a multicellular body of clonal cells, each cell will maximize its (cell level) vehicle quality by playing its part in building a coherent body that, in turn, maximizes its (organism level) vehicle quality. Similarly, if eusocial colonies ('superorganisms') have strong control mechanisms against selfishness to quickly eliminate all incipient 'rogue' traits (thus rendering selfish adaptations effectively impossible), then, like cells in a body, the organisms in a superorganism should become adapted to maximise their vehicle quality through maximizing the superorganism's vehicle quality. However, control mechanisms against 'rogue traits' probably arise more easily at the organism than superorganism level. For example, consider an organismal adaptation that involves a network of genes that interact in a coordinated fashion to produce a coherent outcome. This network operates in a biochemical environment that is readily accessible to gene products from all the other genes of the organism. This accessibility creates thousands of possibilities for mutations to undermine the adaptation. Moreover, the mechanisms doing the undermining could be as simple and low-cost as one protein binding to another, rather than needing to be complex adaptations in their own right. This makes complex 'rogue adaptations' within organisms extremely unlikely. By contrast, because these arguments do not apply to the same extent to superorganisms, there is less reason to expect a priori that adaptations improve vehicle quality at the superorganism level.

*Q39 Does the kinship theory of genomic imprinting, which posits that evolutionary interests may differ between paternally versus maternally-derived genes, contradict the idea that an individual's genome has a 'majority interest'?*
No. Imprinting may cause deviations from an individual's optimal phenotype, but it does not negate the usefulness of defining an optimum. To a first approximation, imprinting can be understood as a manifestation of a conflict of interests between organisms – namely between parents or between parents and offspring – played out within the offspring's genomes [55,56]. In addition, maternally and paternally imprinted genes have interests of their own, favouring (i.e. being best propagated by) phenotypes that differ from the optima of any of the organisms involved [57]. Although the interplay of these interests should cause fluctuations in the evolutionary trajectory of offspring phenotypes, it seems questionable whether the influence of the relatively small number of imprinted loci, which "pull" phenotypic evolution in opposing directions, is anywhere near as strong as that due to the unimprinted majority of genes, in shaping phenotypes through cumulative multi-locus evolution. We might expect that, in general, offspring will exhibit a phenotype (e.g. nutritional demand) close to their own optimum. Note that this view is consistent with the persistence of imprinting even after its phenotypic effects have been eliminated. For example, if a growth factor locus is silent when maternally imprinted yet is highly expressed when paternally imprinted, selection on unimprinted genes may (re-)establish the optimum offspring phenotype by affecting downstream mechanisms activated by the growth factor.

*Q40 Do you have anything to add for readers who are still sceptical?*
A thought experiment might help. Imagine you are a bioengineer in the distant future, when one can change organisms by rewriting their DNA. Assume you are faced with the following task. In your lab you have numerous animal embryos, each of them individually taken from a wild mother from a separate source population. The embryos are sequenced and, by comparison with their source populations, all alleles are identified that match the definition of a reference gene (i.e., that are rarely or never expressed, and rare in the source population). Your task is to modify each embryo's somatic DNA (but not the germline) to create an animal that will increase the average population-wide frequency of its reference genes as much as possible. The modified embryos will then be implanted back into their mothers to continue their development. Essentially, your goal is to design organisms that are good at propagating low-penetrance genes. Your success will be measured by recording future gene frequencies. A number of useful observations follow:

- (i) There is an objective sense in which some organismal designs are better than others to advance your goal.
- (ii) In each population, the focal embryo is the only lever you can pull to affect the target variable.
- (iii) If a focal organism dies early on, that amounts to fewer reference gene copies (i.e., a cost). And if a focal organism helps its sibling to produce > 2 nieces or nephews, at a cost of one of its own offspring, that is a net benefit. And so forth. To optimise traits for their cost-benefit balance, none of the focal organism's offspring should be neglected ('stripped'), because they all have the same potential to contribute to your goal.
- (iv) Designing animals that dramatically outperform wild-type animals won't be easy. Your best bet may be to build a marginally improved animal (e.g. with a few immune genes added to increase survival), or to use a qualitatively different design that could not have evolved gradually. There should be little potential to improve the design by adjusting quantitative traits, because natural selection would already have done so.

(v) Now envisage a design D that outperforms wild-type animals, and that could plausibly have arisen by a natural mutation. How would selection act on a low-penetrance germline mutation that happens to induce D? Answer: it would be positively selected by the same mechanism that D was designed to optimize.

(vi) Hence, only populations in which wild-type animals are optimized to propagate their low-penetrance genes can be phenotypically stable.